\documentclass[epj]{tc}

\usepackage{graphics}
\usepackage{amsmath}
\usepackage{amssymb}
\usepackage{times}
\usepackage{dsfont}
\usepackage{theorem}

\def\sq{\hbox{\rlap{$\sqcap$}$\sqcup$}}
\def\pdh{(2\pi )^{-3/2}}
\def\io{\int{d^3k\over\sqrt{2\omega}}\,}
\def\d{\partial}
\def\=d{\,{\buildrel\rm def\over =}\,}
\def\dl{\dh^\leftrightarrow}
\def\dh{\mathop{\vphantom{\odot}\hbox{$\partial$}}}

\begin{document}

\title{Aspects of the derivative coupling model in four dimensions}
\author{Andreas Aste\inst{1}
\inst{2}
}
\institute{Department of Physics, University of Basel, 4056 Basel, Switzerland \and
Paul Scherrer Institute, 5232 Villigen PSI, Switzerland}
\date{January 23, 2014}
\abstract{
A concise discussion of a 3+1-dimensional derivative coupling model, in which a massive
Dirac field couples to the four-gradient of a massless scalar field, is given in order to elucidate
the role of different concepts in quantum field theory like the regularization of quantum fields as
operator valued distributions, correlation distributions, locality, causality, and field operator gauge
transformations.
\PACS{
      {11.10.-z}{Field theory}   \and
      {11.10.Gh}{Renormalization} \and
      {11.15.-q}{Gauge field theories}
     }
}

\maketitle
\section{Introduction}
\label{intro}
Quantum field theory (QFT) is plagued by many conceptual problems.
It has hitherto been impossible to prove the existence of a non-trivial QFT
in four space-time dimensions. E.g., it is notoriously difficult for perturbative QFTs
to establish convergence of expansions of the $S$-matrix and related observable
quantities. Despite this fact, perturbative QFT has been very successful
in predicting measurable quantities in elementary particle physics. On the perturbative level,
infrared and ultraviolet divergences can be handled by several mathematical tricks and tools.
Whereas ultraviolet divergences are rather related to the short distance behaviour of a
QFT, integrals over infinite space-time result in some sort of infrared difficulties
when massless fields are involved, depending on the approach that was chosen to formulate the theory.\\

\noindent As a general remark, one may say that QFT on unquantized space-time
can be considered as some sort of operator valued distribution theory, which respects basic inputs
coming from symmetry considerations which normally include the Poincar\'e symmetry group
$\mathcal{P}^\uparrow_+$ as the semidirect product of the abelian group of time-space translations $T_{1,3}$ and
the restricted Lorentz group $SO^+(1,3)$, or, to be more precise,  the covering group
$\bar{\mathcal{P}}^\uparrow_+ = T_{1,3} \rtimes SL(2,\mathds{C})$ \cite{PCT}.\\

\noindent Even the definition of a particle in non-gravitating flat space-time becomes a non-trivial task when
charged particles coupling to massless
gauge fields become involved. Based on the classical analysis of Wigner on the unitary representations of
the Poincar\'e group, a one-particle state is an element of an irreducible representation space
of the double cover of the Poincar\'e group in a
physical Hilbert space, i.e. some irreducible representations should occur
in the discrete spectrum of the mass-squared operator $M^2=P_\mu P^\mu$ of a QFT
describing particles \cite{Wigner}. However, objects like the electron are accompanied by a long range field
which leads an independent life at infinite spatial distance, to give an intuitive picture. It has been shown
in \cite{Buchholz} that a discrete eigenvalue of $M^2$ is absent for states with an electric charge as a direct
consequence of Gauss' law, and one finds that the Lorentz symmetry is not implementable in a sector of
states with nonvanishing electric charge. Such problems are related to the fact that the Poincar\'e symmetry
is an overidealization related to global considerations of infinite flat space-time, however, physical measurements have
a local character.\\

\noindent In this paper, we follow a shut up and calculate approach, in order to hint at the fact that
many aspects of QFT are still poorly understood and to demonstrate the mathematical
apparatus which is treated very often on a fairly phenomenological level. The derivative coupling model,
which serves thereby as a trivial, but stunning example for this fact, will be discussed in two different
versions.

\section{The classical derivative coupling model}
As a starting point for the derivative coupling model discussed in this paper, one may consider
the equations of motion of the coupled Maxwell-Dirac system where a massive spin-$1/2$ field $\psi$
couples to a massless abelian spin-$1$ gauge field $A_\mu$ in the Feynman gauge
\begin{equation}
(i \gamma_\mu \partial^\mu -m ) \psi(x) = e A^\mu(x) \gamma_\mu \psi(x) \, ,
\end{equation}
\begin{equation}
\Box A_\mu(x) = j_\mu(x) = e \bar{\psi}(x) \gamma_\mu \psi(x) \, ,
\end{equation}
where, e.g., a coupling constant $e<0$ would relate to a field $\psi$ describing negatively charged objects
like electrons as particles and the positively charged positrons as anti-particles.
$\gamma^0, \ldots , \gamma^3$ are Dirac matrices fulfilling the standard anticommutation relations.
Replacing $A_\mu$ by the four-gradient of a massless, neutral scalar field
 $\varphi$ \cite{Schroer} and, in order to clearly distinguish the two theories from a notational
point of view, the electric coupling constant
 $e$ by a coupling constant $g$ leads to the defining equations of the derivative coupling model
\begin{equation}
(i \gamma_\mu \partial^\mu -m ) \psi(x) = g \partial^\mu \varphi(x) \gamma_\mu \psi(x) \,  , \label{deriv1}
\end{equation}
\begin{equation}
\Box \varphi(x) = \partial^\mu j_\mu(x) = 0 \, . \label{deriv2}
\end{equation}
These equations can be derived from the Lagrangian
\begin{displaymath}
\mathcal{L}= i \bar{\psi} \gamma^\mu \partial_\mu \psi  - m \bar{\psi} \psi +
\frac{1}{2} \partial_\mu \varphi  \partial^\mu \varphi
- g \partial^\mu \varphi \bar{\psi} \gamma_\mu \psi
\end{displaymath}
\begin{equation}
= \mathcal{L}^0_\psi + \mathcal{L}^0_\varphi + \mathcal{L}_{int} \,
\end{equation}
with
\begin{equation}
\mathcal{L}_{int}= - g \partial^\mu \varphi \bar{\psi} \gamma_\mu \psi \, .
\end{equation}

\noindent In \emph{classical} field theory,  a solution of eqns. (\ref{deriv1}) and (\ref{deriv2}) is readily found\\
\begin{equation}
\psi(x)= e^{-ig\varphi(x)} \psi_0(x) \, ,
\end{equation}
with free fields $\varphi(x)$ and $\psi_0(x)$ satisfying
\begin{equation}
\Box \varphi(x)=0, \quad (i \gamma_\mu \partial^\mu -m ) \psi_0(x) =0 \, ,
\end{equation}
since one has
\begin{displaymath}
(i \gamma_\mu \partial^\mu-m) \psi=
i \gamma_\mu \partial^\mu (\psi_0 e^{-ig \varphi})-m \psi
\end{displaymath}
\begin{equation}
= i \gamma_\mu  e^{-ig\varphi(x)}
\partial^\mu \psi_0 -m \psi +g  \partial^\mu
\varphi \gamma_\mu \psi
= g \partial^\mu \varphi  \gamma_\mu \psi \, .
\end{equation}

\noindent
Leaving the classical level, it may be argued that
the interacting Dirac field is 'dressed' in some sense by excitations of the massless bosonic field.
However, since quantum fields are operator valued distributions, products or exponentials of such objects are not defined in
general and require a thorough discussion. Field products are unavoidable for the construction of observables,
since neither the Dirac field nor the vector potential correspond to observable quantities. Still, it seems evident
that the derivative coupling model is physically trivial since the Dirac field couples to a pure gauge. The model itself
is invariant under gauge transformations
\begin{equation}
\psi'(x)= e^{-ig \chi(x)} \psi(x) \, , \quad \varphi'(x)=\varphi(x)+\chi(x) \, ,
\end{equation}
where again $\Box \chi(x)=0$,
and a mass term for the scalar field $\varphi$ could be included in the model, but this option will not
be considered in this paper.

\section{Preliminaries and conventions}
\subsection{The free scalar field}
In order to provide a well-defined setting for the forthcoming discussion of the derivative coupling model
on a quantum field theoretical level, we discuss some basic properties and definitions concerning the free, i.e.
non-interacting scalar field describing a neutral or charged spin-0 particle of mass $M$ in (3+1) space-time dimensions.
Such a discussion may appear as an overkill, but it is not.
Scalar bosonic fields may be represented according to
\begin{displaymath}
\varphi(x)=\varphi^{-}(x)+\varphi^{+}(x)
\end{displaymath}
\begin{equation}
=
\frac{1}{(2 \pi)^{3/2}} \int \frac{d^3 k}{\sqrt{2 k^0}}
[a(\vec{k}) e^{-ikx}+a^\dagger(\vec{k}) e^{+ikx}] \quad \mbox{(neutral)},
\end{equation}
\begin{displaymath}
\varphi_c(x)=\varphi_c^{-}(x)+\varphi_c^{+}(x)
\end{displaymath}
\begin{equation}
=
\frac{1}{(2 \pi)^{3/2}} \int \frac{d^3 k}{\sqrt{2 k^0}}
[a(\vec{k}) e^{-ikx}+b^\dagger(\vec{k}) e^{+ikx}] \quad \mbox{(charged)},
\end{equation}
where $k x= k_\mu x^\mu=k^0 x^0 - \vec{k} \cdot \vec{x}$, $k^0 \overset{!}{=}E=
\sqrt{\vec{k}^2+M^2}>0$,
$\pm$ denotes the positive and negative frequency parts of the fields and
$\dagger$ a 'hermitian conjugation' .
The non-vanishing \emph{distributional} commutator relations for the destruction and creation field operators
in the above Fourier decomposition are
\begin{equation}
[a(\vec{k}),a^\dagger(\vec{k'})]=[b(\vec{k}),b^\dagger(\vec{k'})]=
\delta^{(3)} (\vec{k}-\vec{k'}) \, , \label{algeb1}
\end{equation}
otherwise
\begin{displaymath}
[a(\vec{k}),a(\vec{k'})]=[b(\vec{k}),b(\vec{k'})]
\end{displaymath}
\begin{equation}
=[a^\dagger
(\vec{k}),a^\dagger(\vec{k'})]
=[b^\dagger(\vec{k}),b^\dagger(\vec{k'})]= 0  \label{algeb2}
\end{equation}
and
\begin{displaymath}
[a(\vec{k}),b(\vec{k'})]=[a(\vec{k}),b^\dagger(\vec{k'})]
\end{displaymath}
\begin{equation}
=[a^\dagger
(\vec{k}),b(\vec{k'})]=[a^\dagger(\vec{k}),b^\dagger(\vec{k'})]= 0  \label{algeb2a}
\end{equation}
holds.
The destruction (or 'annihilation', or 'absorption')  operators act on the \emph{non-degenerate} vacuum
$|0\rangle$ according to
\begin{equation}
a(\vec{k}) |0 \rangle=b(\vec{k}) |0 \rangle = 0 \quad \mbox{for all}  \, \,  k \in \mathds{R}^3 \, . \label{vacstate}
\end{equation}
It is crucial to require the existence of a state $| 0 \rangle$ which is annihilated by
all the $a(\vec{k})$ and  $b(\vec{k})$, since otherwise there would be many inequivalent irreducible Hilbert space
representations of the algebraic relations given by eqns. (\ref{algeb1}) -(\ref{algeb2a}), and eq. (\ref{vacstate})
selects the one in Fock space where the $a(\vec{k})$ and  $b(\vec{k})$ can be interpreted as destruction and
the $a^\dagger(\vec{k})$ and  $b^\dagger(\vec{k})$ as creation (or 'emission') operators.\\

\noindent Single-particle wave functions in momentum space $\Psi_1(\vec{k})$, $\Psi_2(\vec{k})$
are
\begin{displaymath}
| \Psi_1  \rangle = \int d^3 k \, \Psi_1(\vec{k}) a^\dagger(\vec{k})|0 \rangle  \, ,
\end{displaymath}
\begin{equation}
| \Psi_2 \rangle = \int d^3 k' \, \Psi_2(\vec{k'}) a^\dagger(\vec{k'}) |0 \rangle \, , \label{single_particle}
\end{equation}
their scalar product becomes from a formal calculation exploiting the commutation relations above
\begin{displaymath}
\langle \Psi_1 | \Psi_2 \rangle =
\int d^3 k  d^3 k' \, {\Psi_1}(\vec{k})^* \Psi_2(\vec{k'})\langle 0 | a(\vec{k}) a^\dagger(\vec{k'}) 0 \rangle
\end{displaymath}
\begin{displaymath}
=\int d^3 k  d^3 k' \, {\Psi_1}(\vec{k})^* \Psi_2(\vec{k'}) \langle 0 | [\delta^{(3)}(\vec{k}-\vec{k'}) +
a^\dagger(\vec{k'}) a(\vec{k}) ] | 0 \rangle
\end{displaymath}
\begin{equation}
=
\int d^3 k  \, {\Psi_1}(\vec{k})^* \Psi_2(\vec{k})  \, .
\end{equation}
This scalar product can be written in a manifestly covariant form by using differently normalized
creation and destruction operators fulfilling
\begin{equation}
[a(\vec{k}),a^\dagger(\vec{k'})]=[b(\vec{k}),b^\dagger(\vec{k'})]=
(2 \pi)^{3/2} (2 k^0)^{1/2} \delta^{(3)} (\vec{k}-\vec{k'}).
\end{equation}

\subsection{Quantum fields as operator valued distributions}
\noindent It is crucial to note that $\varphi(x)$ and $\varphi_c(x)$ are \emph{operator valued distributions}, i.e.
only smeared out fields like
\begin{equation}
\varphi(g)=\int d^4 x \, \varphi(x) g(x) \, , \label{smeared_field}
\end{equation}
where $g$ is a test function is some suitable test function space $\mathcal{T}(\mathds{R}^4)$,
are operators in the quantum mechanical sense on the Hilbert-Fock space of free particles, i.e.
linear operators defined on a dense subset of the Hilbert space which are not necessarily bounded
\cite{Wintner}, \cite{Wielandt}. The same observation applies in momentum space, i.e.
\begin{equation}
a^\dagger(\hat{g})=\int d^4 k \, a^\dagger(k) \hat{g}(k) \, , \label{smeared_a}
\end{equation}
creates a physical, i.e. normalizable Fock state, whereas $a^\dagger(\vec{k})| 0 \rangle$ is not
a vector in Fock space, since no finite norm can be assigned to such an object due to
eq. (\ref{algeb1}).
In fact, smearing field operators of a four-dimensional field theory in three dimensions as anticipated
in eq. (\ref{single_particle}) does not work in general in the case of interacting fields.\\

\noindent It is common usage in QFT in $n$ space-time dimensions
to work with test functions which are elements of
the Schwartz space of rapidly decreasing functions $\mathcal{S}(\mathds{R}^n)$. This space is
obtained by considering complex valued $p$-times continuously differentiable functions in $C^p(\mathds{R}^n)$
equipped with the norms
\begin{displaymath}
||f||_p=\sup_{|\alpha| \le p} \sup_{x \in \mathds{R}^n} (1+||x||)^p |D^{\alpha} f(x)| \ ,
\end{displaymath}
\begin{equation}
x=(x^1, \ldots x^n) \quad, \quad ||x||=\sqrt{\sum_{i=1}^{n} (x^{i})^2} \, ,
\end{equation}
with multiindices $\alpha=(\alpha_1, \ldots \alpha_n) \in \mathds{N}_0^n$ and differential operators
\begin{equation}
D^\alpha = \frac{\partial^{\alpha_1}}{\partial x^{\alpha_1}} \ldots \frac{\partial^{\alpha_n}}{\partial x^{\alpha_n}} \, ,
\, \, \mbox{where} \, \, |\alpha |=\alpha_1+ \ldots \alpha_n \, ,
\end{equation}
defining thereby complete normed function spaces
\begin{equation}
\bar{S}_p(\mathds{R}^n) =\{ f \in C^p(\mathds{R}^n) |  \, \,
||f||_p < \infty  \} \, .
\end{equation}
The Schwartz space $\mathcal{S}(\mathds{R}^n)$ is then defined as the space  of infinitely differentiable
functions of rapid decrease
\begin{equation}
\mathcal{S}(\mathds{R}^n) = \bigcap \limits_{p=0}^{\infty} \bar{S}_p (\mathds{R}^n)  \, .
\end{equation}
By a meaningful definition, a series of test functions
$\{f_\nu\}_{\nu=0}^{\infty} \subset \mathcal{S}(\mathds{R}^n)$ converges towards  $f=0$ iff
$||f_\nu||_p \overset{\nu \rightarrow \infty}{\rightarrow} 0$ for all $p \in
\mathds{N}_0.$
A typical example for a test function in $\mathcal{S}(\mathds{R})$ is given by
$g(x)= e^{-x^2}$.\\

\noindent The space of  \emph{tempered distributions} $\mathcal{S}'(\mathds{R}^n)$ is the set of
the continuous linear functionals on  $\mathcal{S}(\mathds{R}^n)$ according to
\begin{equation}
d \in \mathcal{S}'(\mathds{R}^n)  \, \, \Leftrightarrow \, \, d(f_\nu) \rightarrow 0 \, \, \mbox{for all}
 \, \{f_\nu\}_{n=0}^{\infty} \subset
\mathcal{S}(\mathds{R}^n)
\end{equation}
where $ f_\nu \overset{\nu \rightarrow \infty}{\rightarrow} 0$.
This definition of a tempered distribution becomes more intuitive if one realizes that
such an object can be represented as the sum of derivatives of continuous functions of polynomial growth
\begin{displaymath}
d \in \mathcal{S}(\mathds{R}^n) \, \,  \Leftrightarrow \, \, d(f)
\end{displaymath}
\begin{displaymath}
= \sum \limits_{0 \le |\alpha | \le s \in \mathds{N}}
\int dx^1 \ldots dx^n
\end{displaymath}
\begin{equation}
\times (-1)^{|\alpha|} F_\alpha (x^1, \ldots x^n) D^\alpha f(x^1, \ldots x^n)  \, ,
\end{equation}
where
$C(\mathds{R}^n) \ni F_\alpha (x)$, $|F_\alpha (x) | \le c_\alpha (1+ ||x||)^{j(\alpha)}$ for some $j(\alpha) \in \mathds{N}$
and $c_\alpha \in \mathds{R} \, .$
Formally, derivatives can be shifted by partial integration from test functions to distributions.\\

\noindent The true reason for using the Schwartz space in QFT is its convenient property that
the Fourier transform acts on $\mathcal{S}(\mathds{R}^n)$  as a unitary, bijective mapping, i.e the
Fourier transform of a smooth, rapidly decreasing function is again smooth and rapidly decreasing.
This allows to define the Fourier transform $\mathcal{F}$
of singular objects like the distributions in $\mathcal{S}'(\mathds{R}^n)$.
$\hat{d}=\mathcal{F}(d)$ is defined so that
for all $ f \in \mathcal{S}(\mathds{R}^n)$

\begin{equation}
\mathcal{F}(d)(f) = \hat{d}(f) = d(\mathcal{F}(f)) = d(\hat{f})  \, ,
\end{equation}
a definition which is often expressed by the purely formal expression involving a change in the
order of integration
\begin{displaymath}
\hat{d}(f) = \int \limits_{\mathds{R}^n} dx \,  \hat{d}(x) f(x)
\end{displaymath}
\begin{equation}
= \frac{1}{(2 \pi)^{n/2}}
\int \limits_{\mathds{R}^n} dx \,  \int  \limits_{\mathds{R}^n}  dk \,
d(k) e^{-ikx}  f(x)
= \int  \limits_{\mathds{R}^n} dk \, d(k) \hat{f}(k) \, .
\end{equation}
Equivalently we have
\begin{equation}
\hat{d}(\check{f}) = d(\hat{\check{f}}) = d(f) \, .
\end{equation}

\noindent This way, the Fourier transform also becomes a linear automorphism of $\mathcal{S}'$\\
\begin{equation}
\mathcal{F}(\mathcal{S}(\mathds{R}^n)) = \mathcal{S}(\mathds{R}^n) \, , \quad
\mathcal{F}(\mathcal{S}'(\mathds{R}^n)) = \mathcal{S}'(\mathds{R}^n) \, .
\end{equation}

\noindent Throughout this paper, the Fourier transform of a function
on four-dimensional space-time will be defined according to the sign
and symmetric normalization convention
\begin{displaymath}
\hat{\Phi}(k)=\frac{1}{(2 \pi)^2} \int  \limits_{\mathds{R}^4} d^4 x \,\Phi(x) e^{ikx}
\end{displaymath}
\begin{equation}
= \frac{1}{(2 \pi)^2} \int  \limits_{\mathds{R}^4}d^4 x \, \Phi(x) e^{ik^0x^0- i \vec{k} \vec{x}} \, ,
\end{equation}
with $kx=k_ \mu x^\mu = k_0 x^0+k_1 x^1 + k_2 x^2 + k_3 x^3=k^0 x^0-k^1 x^1 - k^2 x^2 -k^3 x^3$.\\

\noindent An important subspace of distributions in $\mathcal{D}(\mathds{R}^n) \subset \mathcal{S}(\mathds{R}^n)$
is spanned by the distributions of compact support. The dual space $\mathcal{D}'(\mathds{R}^n)$
of linear functionals on this space is more general than $\mathcal{S}'(\mathds{R}^n)$ and contains it.
For the sake of brevity, topological aspects of $\mathcal{D}(\mathds{R}^n)$ and $\mathcal{D}'(\mathds{R}^n)$ will
not be discussed here. However, it is important to note that causality in QFT is often expressed by a
relation of the form
\begin{equation}
[O_1(g_1),O_2(g_2)] = 0 \quad \mbox{for} \, \, \mbox{supp}(g_1)  \sim \mbox{supp}(g_2) \, ,
\label{causality}
\end{equation}
which expresses the fact that two local observables $O_1$ and $O_2$ depending as operator valued
distributions on test functions $g_1$, $g_2 \in \mathcal{D}(\mathds{R}^n) \subset \mathcal{S}(\mathds{R}^n)$
commute whenever the compact supports of the test functions
are space-like separated, i.e. when $(x_1-x_2)^2 < 0$ holds for all $x_1 \in \mbox{supp}(g_1)$
and $x_2 \in \mbox{supp}(g_2)$. One should note that the Fourier transforms $\hat{g_1}$ and
$\hat{g_2}$ do \emph{not} have compact support for $g_1$, $g_2 \neq 0$. The commutator
eq. (\ref{causality}) may become an anticommutator when fermionic fields are involved. However, such
fields are elements of a field algebra and not of an algebra of observables, but they often serve as
building blocks for the construction of observables.\\

\noindent In Appendix A, a well-know but indispensable set of relations needed for the manipulation of
distributions is given for the reader who only has enjoyed a cursory formal introduction to the theory.

\subsection{Correlation distributions}
\noindent From the above algebraic relations represented by free fields on a Fock space $\mathcal{F}$
one constructs the scalar Feynman propagator as  distributional time-ordered vacuum expectation values
\begin{equation}
\Delta_F(x-y)=-i \langle 0 | T (\varphi_c (x) \varphi_c^\dagger(y) | 0 \rangle \, ,
\end{equation}
where translational invariance implies
\begin{equation}
\Delta_F(x)=-i \langle 0 | T (\varphi_c (x) \varphi_c^\dagger(0) | 0 \rangle
\end{equation}
or
\begin{equation}
\Delta_F(x)=-i \langle 0 | T (\varphi(x) \varphi(0) | 0 \rangle,
\end{equation}
for neutral fields.
The wave equation holds in a distributional sense
\begin{equation}
(\Box+M^2) \Delta_F(x)=(\partial_\mu \partial^\mu+M^2) \Delta_F(x)=- \delta^{(4)} (x)
\end{equation}
and one also defines the positive- and negative-frequency Pauli-Jordan $C$-number distributions
or, up to an imaginary factor, 'Wightman two-point functions'
\begin{equation}
\Delta^{\pm}(x)=-i[\varphi^{\mp}(x),\varphi^{\pm}(0)] =-i[\varphi_c^{\mp}(x),\varphi_c^{\dagger \, \pm}(0)]  \, ,
\end{equation}
\begin{displaymath}
\Delta(x)=\Delta^{+}(x)+\Delta^{-}(x)
\end{displaymath}
\begin{equation}
= -i [\varphi(x),\varphi(0)] = -i [\varphi_c(x), \varphi_c^\dagger(0)]\, ,
\end{equation}
i.e.
\begin{displaymath}
\Delta^{+}(x)=-i \langle 0 | \varphi^- (x) \varphi^+ (0) | 0 \rangle \, ,
\end{displaymath}
\begin{equation}
\Delta^{-}(x)=+i \langle 0 | \varphi^- (0) \varphi^+ (x) | 0 \rangle \, . \label{delta_plus}
\end{equation}
The retarded propagator is given by $\Delta^{ret}(x)=\Theta(x^0) \Delta(x)$, a product of distributions
which is well-defined due to the harmless scaling behaviour of $ \Delta(x)$ at the origin $x=0$.\\

\noindent Some important properties of the objects and their Fourier transforms introduced so far are enlisted
in the following:
$\Delta(x)$ vanishes for space-like arguments $x$ with $x^2 < 0$, as required by causality. One has
\begin{displaymath}
\hat{\Delta}^{\pm}(k)=\frac{1}{(2 \pi)^2} \int d^4 x \, \Delta^{\pm} (x) e^{ikx}
\end{displaymath}
\begin{equation}
=\mp \frac{i}{2 \pi} \Theta(\pm k^0)
\delta(k^2-M^2) \, ,
\end{equation}
\begin{equation}
\Delta^+(x) = - \Delta^-(-x) \, \ ,
\end{equation}
\begin{equation}
\Delta(x)=\Delta^+(x)-\Delta^+(-x) \, ,
\end{equation}
\begin{equation}
\Delta(-x)=-\Delta(x) \, .
\end{equation}
\begin{equation}
\Delta_F(x)=\Theta(x^0) \Delta^+(x) - \Theta(-x^0) \Delta^-(x) \, .
\end{equation}
\begin{equation}
(\Box+M^2) \Delta^{\pm}(x)=0 \, ,  \quad (k^2-M^2) \hat{\Delta}^\pm(k)=0 \, .
\end{equation}
\begin{equation}
\Delta^{ret}=\Delta_F+\Delta^{-} \, ,
\end{equation}
\begin{equation}
\displaystyle \Delta^{ret}(x)=\int \frac{d^4 k}{(2 \pi)^4} \frac{e^{-ikx}}{k^2-M^2+i k^0 0} \, ,
\end{equation}
\begin{equation}
(\Box+M^2) \Delta^{ret}(x)=- \delta^{(4)} (x) \, .
\end{equation}
For $M=0$ the scalar Feynman propagator in configuration space is
\begin{displaymath}
\Delta_F^0(x)
=\int \frac{d^4 k}{(2 \pi)^4} \frac{e^{-ikx}}{k^2+i0}
\end{displaymath}
\begin{equation}
=\frac{i}{4 \pi^2} \frac{1}{x^2-i0} =
\frac{i}{4 \pi^2} P \frac{1}{x^2}-
\frac{1}{4 \pi} \delta (x^2) \,  ,
\end{equation}
where $P$ denotes the principal value and  $\delta$ the one-dimensional  Dirac distribution, and
the massless Pauli-Jordan distributions in configuration space are
\begin{equation}
\displaystyle \Delta^0(x)
=-\frac{1}{2 \pi} \mbox{sgn} (x^0) \delta(x^2) \,  ,
\end{equation}
\begin{equation}
\displaystyle  \Delta^\pm_0(x) = \pm \frac{i}{4 \pi^2} \frac{1}{(x_0 \mp i0)^2 -\vec{x}^2} \, ,
\end{equation}
and since $\Delta^{ret}(x)=\Theta(x^0) \Delta(x)$ one has
\begin{equation}
\Delta^{ret}_0(x)
=-\frac{1}{2 \pi} \Theta (x^0) \delta(x^2) \, .
\end{equation}

\noindent A notational issue concerning the principal value in the case of  $\Delta_0^+$ is clarified by
\begin{displaymath}
\frac{1}{(x^0-i0)^2-\vec{x}^2}=\frac{1}{((x^0-i0)-|\vec{x}|)((x^0-i0)+|\vec{x}|)}
\end{displaymath}
\begin{displaymath}
=\frac{1}{2|\vec{x}|} \frac{1}{x^0-|\vec{x}|-i0}- \frac{1}{2|\vec{x}|} \frac{1}{x^0+|\vec{x}|-i0}
\end{displaymath}
\begin{equation}
=P \frac{1}{x^2}+i \pi \mbox{sgn}(x^0) \delta(x^2)
\end{equation}
or
\begin{displaymath}
\frac{1}{(x^0-i0)^2-\vec{x}^2}=
\frac{1}{x^2-2i0 x^0-0^2}=\frac{1}{x^2-i0 \mbox{sgn}(x^0)}
\end{displaymath}
\begin{equation}
= P \frac{1}{x^2}+i \pi \mbox{sgn}(x^0) \delta(x^2) \, .
\end{equation}

\subsection{Positivity}
\noindent Calculating explicitly the commutator
\begin{displaymath}
 [\varphi_c(x), \varphi_c^\dagger(y)] =
\frac{1}{(2 \pi)^3} \int \frac{d^3k'}{\sqrt{2E'}}  \int \frac{d^3k}{\sqrt{2E}}
\end{displaymath}
\begin{displaymath}
[ a(\vec{k}') e^{-ik'x} + b^\dagger(\vec{k}')e^{+ik'x},
a^\dagger(\vec{k}) e^{+iky} + b(\vec{k})e^{-iky}] =
\end{displaymath}
\begin{displaymath}
\frac{1}{(2 \pi)^3} \int \frac{d^3k'}{\sqrt{2E'}}  \int \frac{d^3k}{\sqrt{2E}}
\end{displaymath}
\begin{displaymath}
[ \delta^{(3)}(\vec{k}'-\vec{k}) e^{-ik'x+iky}
-\delta^{(3)}(\vec{k}'-\vec{k}) e^{+ik'x-iky)}]=
\end{displaymath}
\begin{displaymath}
\frac{1}{(2 \pi)^3} \int \frac{d^3 k}{2 E} \{ e^{-ik(x-y)}-e^{+ik(x-y)} \}=
\end{displaymath}
\begin{equation}
\frac{1}{(2 \pi)^3} \int d^4k \, \mbox{sgn}(k^0) \delta(k^2-M^2)  e^{-ik(x-y)} \, ,
\end{equation}
where
\begin{displaymath}
 \mbox{sgn}(k^0) \delta(k^2-M^2) =  \mbox{sgn}(k^0) \delta(k_0^2 -\vec{k}^2 - M^2)
\end{displaymath}
\begin{equation}
=
\frac{ \mbox{sgn}(k^0)}{2 |k^0|} \{ \delta(k^0-E) + \delta(k^0+E) \}
\end{equation}
has been used, one finds one of the results given above
\begin{equation}
\hat{\Delta}(k)= -\frac{i}{2 \pi} \mbox{sgn}(k^0) \delta(k^2-M^2) \, . \label{positivity}
\end{equation}
At the same time, at glimpse at the calculation above reveals
\begin{displaymath}
\hat{\Delta}^{+}(k)=\frac{1}{(2 \pi)^2} \int d^4 x \, \Delta^{+} (x) e^{ikx}
\end{displaymath}
\begin{equation}
=- \frac{i}{2 \pi} \Theta(+k^0)
\delta(k^2-M^2) \, . \label{positivity2}
\end{equation}

\noindent Eq. (\ref{positivity}) simply expresses the fact that the scalar fields considered so far live
in a Hilbert space, equipped by definition with a positive definite norm.
Indeed, creating a one-particle state by acting with a smeared field operator on the vacuum
\begin{equation}
| \Phi \rangle = \int d^4 x \, \Phi(x) \varphi(x) | 0 \rangle =
\int d^4 x \, \Phi(x) \varphi^+(x) | 0 \rangle
\end{equation}
and calculating the norm gives, using eq. (\ref{delta_plus})
\begin{displaymath}
\langle \Phi | \Phi \rangle =  i \int d^4 x' \, d^4 x \, \Phi(x')^* \Delta^+(x'-x) \Phi(x)=
\end{displaymath}
\begin{displaymath}
\frac{i}{(2 \pi)^6} \int d^4 k'' \, d^4 k' \, d^4 k \, d^4 x' \, d^4 x \,
\end{displaymath}
\begin{equation}
\hat{\Phi}(-k')^* e^{-i k' x'} \hat{\Delta}^+ (k'') e^{-i k'' (x'-x)} \hat{\Phi} (k) e^{-ikx} \, ,
\end{equation}
where the non-vanishing test function and the positive-frequency Pauli-Jordan distribution have been replaced
their corresponding Fourier transforms. Using the distributional identity
\begin{equation}
\int d^4 k \, e^{+ikx} = (2 \pi)^4 \delta^{(4)} (x)
\end{equation}
is allowed here and leads to
\begin{displaymath}
\langle \Phi | \Phi \rangle = i (2 \pi)^2  \int d^4 k'' \, d^4 k' \, d^4 k  \,
\end{displaymath}
\begin{displaymath}
\hat{\Phi}(-k')^* \hat{\Delta}^+ (k'')  \hat{\Phi} (k) \delta^{(4)} (k'+k'') \delta^{(4)}(k-k'')
\end{displaymath}
\begin{displaymath}
=i (2 \pi)^2 \int d^4 k'' \, \hat{\Phi}(k'')^* \hat{\Delta}^+ (k'')  \hat{\Phi} (k'')
\end{displaymath}
\begin{equation}
 =2 \pi \int d^4 k \,  \Theta(+k^0) \delta(k^2-M^2)
\hat{\Phi}(k)^*  \hat{\Phi} (k) >0  \, ,
\end{equation}
i.e., the Heaviside- and $\delta$-distributions in eq. (\ref{positivity2}) express
the fact that states created by bosonic scalar field operators
have positive norm.\\

\noindent We will see below that the derivative coupling model can also be quantized by using fermionic scalar fields,
i.e. ghosts, which exhibit some properties invoking some conceptual differences to the discussion above.


\section{The derivative coupling model: Bosonic version}
\subsection{General considerations}
\noindent The transition from the classical derivative coupling model according to eqns. (\ref{deriv1}) and
(\ref{deriv2}) to a quantized version generates a problem. The exponential\\
\begin{equation}
e^{-ig\varphi(x)}= \sum_{n=0}^{\infty} \frac{(-ig \varphi(x))^n}{n!}
\end{equation}
is not well-defined as an operator valued distribution, since already $\varphi(x) \varphi(x)$ is \emph{ill-defined}.
E.g., a short calculation shows that $\langle 0 | \varphi(x) \varphi(x) | 0 \rangle$ is a divergent expression
which has to be regularized.
A way out of this situation is offered by the normal ordering of field operators which corresponds to a
recursive point-splitting regularization
\begin{equation}
:\varphi(x): = \varphi(x) \,  ,
\end{equation}
\begin{equation}
:\varphi(x)^2: =\lim_{y \rightarrow x} [ \varphi(x) \varphi(y) - \langle 0 | \varphi(x) \varphi(y) | 0 \rangle ] \,  ,
\end{equation}
\begin{displaymath}
:\varphi(x)^n: =\lim_{y \rightarrow x} [: \varphi(x)^{n-1}: \varphi(y)
\end{displaymath}
\begin{equation}
- (n-1) \langle 0 | \varphi(x) \varphi(y) | 0 \rangle :\varphi(x)^{n-2}: ] \,  .
\end{equation}

\noindent The normally ordered product $:\varphi(x)^n:$ is an operator-valued distributions, as well as the tensor product
$:\varphi(x)^n:$$:\varphi(y)^n:$ \cite{Constantinescu}.\\

\noindent Literally, normal ordering products of free field operators
moves all destruction operators to the right, so that creation operators are moved
to the left. E.g.,
\begin{displaymath}
\varphi(x) \varphi(y) = ( \varphi^{-}(x) + \varphi^+(x)) ( \varphi^{-}(y) + \varphi^+(y))
\end{displaymath}
\begin{displaymath}
= \varphi^-(x) \varphi^-(y)  + \varphi^+(x) \varphi^+(y) + \varphi^+(x) \varphi^-(y)+ \varphi^+(y) \varphi^-(x)
\end{displaymath}
\begin{displaymath}
 + [\varphi^-(x), \varphi^+(y)]
\end{displaymath}
\begin{equation}
=:\varphi(x) \varphi(y): + i \Delta^+(x-y) \, .
\end{equation}

\noindent Calculating the following vacuum expectation value according to Wick's theorem
\begin{equation}
{\langle 0| \! :\varphi(x)^n::\varphi(0)^n: \! | 0 \rangle}=i^n n! \Delta^+(x)^n \, ,
\end{equation}
is a well-defined procedure, and the expressions
\begin{equation}
 \frac{(-ig)^n :\varphi(x)^n:}{n!} \, ,
\end{equation}
are well-defined composite field operators. But still, the sum
\begin{equation}
:e^{-ig\varphi(x)}:= \sum_{n=0}^{\infty} \frac{(-ig)^n :\varphi(x)^n:}{n!}=
e^{-ig\varphi^+(x)} e^{-ig\varphi^-(x)}
\end{equation}
turns out to be 'harmless' only in $1+1$ dimensions. For the sake of completeness, some basic facts concerning
the derivative coupling model in two space-time dimensions as discussed by Schroer \cite{Schroer} are
provided in the following.

\subsection{The derivative coupling model in two dimensions}
\noindent In $1+1$ dimensions, the neutral scalar field
\begin{displaymath}
\varphi(x)=\varphi^{-}(x)+\varphi^{+}(x)
\end{displaymath}
\begin{equation}
=\frac{1}{\sqrt{2 \pi}} \int \frac{d k^1}{\sqrt{2 k^0}}
[a(\vec{k}) e^{-ikx}+a^\dagger(\vec{k}) e^{+ikx}]
\end{equation}
leads to the two-dimensional positive frequency Pauli-Jordan distribution
\begin{displaymath}
\Delta^+(x-y)=-i \langle 0 | \varphi(x) \varphi(y) | 0 \rangle
\end{displaymath}
\begin{displaymath}
= - \frac{i}{2 \pi} \int d^2 k \, \Theta(k^0) \delta(k^2-M^2) e^{-ik(x-y)}
\end{displaymath}
\begin{displaymath}
=- \frac{i}{2 \pi} \int \frac{d k^1}{2 k^0} e^{-ik(x-y)}
\end{displaymath}
\begin{equation}
= -\frac{i}{2 \pi} K_0 \Bigl(M \sqrt{-(x-y)^2+i (x^0-y^0)0}\Bigr) \, .
\end{equation}

\noindent This integral diverges for $M \rightarrow 0$, since the  modified Bessel function
(or MacDonald function) behaves for $0 <x \ll 1$ like
\begin{equation}
K_0(x) \simeq - \ln \Bigl( \frac{x}{2} \Bigr) - \gamma \, ,
\end{equation}
where $\gamma$ denotes the Euler-Mascheroni constant.
Regularizing in the infrared according to
\begin{equation}
\Delta^+(x; \lambda)=-\frac{i}{2 \pi} \int  \frac{d k^1}{2 |k^1|} \Theta(|k^1|-\lambda) e^{-ikx}
\end{equation}
leads to ($0<\lambda \ll 1$)
\begin{equation}
\Delta^+(x; \lambda) \simeq \frac{i}{4 \pi} \ln(-\mu^2 x^2 + i x^0 0) = \frac{i}{4 \pi} \ln(-x^2 + i x^0 0) +C
\end{equation}
with $\mu=e^\gamma \lambda \, .$

\noindent On the restricted space of test functions
\begin{equation}
\mathcal{K}= \{ f(x) \in \mathcal{S}(\mathds{R}^2) \, | \int d^2 x \, f(x) =0 \}
\end{equation}
the massless field $\varphi(x)$ is an operator valued distribution, as well as\\
\begin{equation}
\Delta^+_{reg}(x) = \frac{i}{4 \pi} \ln(-x^2 + i x^0 0) \, .
\end{equation}

\noindent Therefore, one has
\begin{displaymath}
\langle 0 | : e^{-ig \varphi(x)} : : e^{+ig \varphi(y)} : | 0 \rangle
\end{displaymath}
\begin{equation}
= \sum \limits_{n=0}^{\infty}
\frac{i^n}{n!} (g^2)^n [\Delta_{reg}^+(x-y)]^n =
e^{i g^2 \Delta^+_{reg} (x-y)}
\end{equation}
and
\begin{displaymath}
\langle 0 | \psi(x) \bar{\psi}(y) | 0 \rangle
\end{displaymath}
\begin{displaymath}
= \langle 0 | \psi_0(x) \bar{\psi}_0(y) | 0 \rangle
\langle 0 | : e^{-ig \varphi(x)} : : e^{+ig \varphi(y)} : | 0 \rangle
\end{displaymath}
\begin{displaymath}
 =\langle 0 | \psi_0(x) \bar{\psi}_0(y) | 0 \rangle e^{-\frac{g^2}{4 \pi} \ln(-(x-y)^2+i (x^0-y^0) 0)}
\end{displaymath}
\begin{equation}
 =\langle 0 | \psi_0(x) \bar{\psi}_0(y) | 0 \rangle
 \biggl( \frac{1}{-(x-y)^2+i(x^0-y^0) 0} \biggr)^{g^2/4 \pi} \, ,
\end{equation}
where $\psi_0$ denotes the free fermionic field in two space-time dimensions.
A straightforward calculation \cite{Schroer} also shows that
\begin{equation}
\Delta_F^\psi(k)=\langle 0 | T(\psi(x) \bar{\psi}(y) | 0 \rangle \sim (k^2-m^2)^{g^2/2 \pi -1} \, .
\end{equation}
No meromorphic pole structure appears for $g \ne 0$, although the $S$-matrix of the theory is trivial.
Due to this reason, Schroer coined the expression \emph{infraparticle} for the states described
by the dressed field $\psi(x)$.

\subsection{Four-dimensional aspects}
\noindent In 3+1 dimensions one has
$\displaystyle \Delta^+_0(x) =  \frac{i}{4 \pi^2} \frac{1}{(x_0 - i0)^2 -\vec{x}^2}$, and
\begin{displaymath}
D^+(x-y)=\langle 0 | : e^{-ig \varphi(x)} : : e^{+ig \varphi(y)} : | 0 \rangle
\end{displaymath}
\begin{displaymath}
= \sum \limits_{n=0}^{\infty}
\frac{i^n}{n!} (g^2)^n [\Delta^+_0(x-y)]^n
\end{displaymath}
\begin{equation}
= e^{i g^2 \Delta^+_0 (x-y)}
=  \displaystyle \exp \biggl( {-\frac{g^2}{4 \pi^2 ((x_0 - i0)^2 -\vec{x}^2)}} \biggr)
\end{equation}
is a highly ultraviolet-divergent (still formal) expression as can be anticipated from the singular behaviour in
configuration space for $x \rightarrow 0$. In fact, the exponential of a
free scalar field operator in four space-time dimensions is no longer an operator valued distribution defined on
$\mathcal{S}(\mathds{R}^4)$.\\

\noindent In conventional regularization theory, one would regularize the exponential of a scalar field according to
\begin{displaymath}
e^{-ig \varphi(x)} \, \rightarrow \,
e^{-\frac{i}{2}g^2 \Delta^+_\Lambda (0)} e^{-ig \varphi_\Lambda(x)}
\,  \, \overset{\Lambda \rightarrow \infty}{\rightarrow} \, \,  :e^{-ig \varphi(x)}: \,
\end{displaymath}
\begin{equation}
= \lim \limits_{\Lambda \rightarrow 0} :e^{-ig \varphi_\Lambda(x)}:  \, =
e^{-ig\varphi_\Lambda^+(x)} e^{-ig\varphi_\Lambda^-(x)}
\, ,
\end{equation}
with a scalar field $\varphi_\Lambda (x)$ with ultraviolet-cutoff $\Lambda$ generating a regular
two-point distribution $\Delta^+_\Lambda(x)$.
The field $\psi_{un,\Lambda} (x) = \psi_0 e^{-ig \varphi_\Lambda(x)}$ would not converge to a well-defined
operator valued distribution in any sense.
However, one can write for the renormalized field with ultraviolet cutoff
\begin{displaymath}
\psi_{ren,\Lambda}(x)=:e^{-ig \varphi_\Lambda(x)}: \psi_0(x)=
e^{-\frac{i}{2} g^2 \Delta_\Lambda^+ (0)} \psi_{un,\Lambda} (x)
\end{displaymath}
\begin{equation}
= Z_\Lambda^{-1/2} \psi_{un,\Lambda} (x) \, .
\end{equation}
In the limit $\Lambda \rightarrow \infty$, with $\psi_{un}$ as the unrenormalized formal limit of $\psi_\Lambda$,
one has formally
\begin{equation}
\psi_{ren}(x) = \lim_{\Lambda \rightarrow \infty} Z_\Lambda^{-1/2} \psi_{un,\Lambda}=Z^{-1/2} \psi_{un}(x) \, ,
\end{equation}
where
\begin{equation}
Z_\Lambda^{-1/2} = e^{-\frac{i}{2} g^2 \Delta^+_\Lambda (0)} \, .
\end{equation}
Then
\begin{equation}
\{ \psi_{ren,\alpha} (x^0,\vec{x}) , \bar{\psi}_{ren,\beta} (x^0,\vec{y}) \} =
Z^{-1} (\gamma^0)_{\alpha \beta} \,  \delta^{(3)}(\vec{x}-\vec{y}) \, ,
\label{etcr}
\end{equation}
i. e. the standard equal time anti-commutation relations cannot be fulfilled by the renormalized fields
since $Z \rightarrow \infty$, but the renormalized field $\psi_{ren}$ has well-defined correlation functions.
The distribution $e^{ig^2 \Delta^+(x-y)}$ cannot be restricted to equal times $x^0=y^0$, a non-canonical
property which one expects for interacting fields.\\

\noindent Still, perturbative terms like $\Delta_0^+(x)^n$ can be defined without problems.
In the following, the product in configuration space $\Delta_0^+(x)^2$ is investigated in detail
in configuration as well as in momentum space.
Defining $\Delta_2^+(x)=(\Delta_0^+(x))^2$, one calculates
\begin{displaymath}
\mathcal{F}(\Delta_0^+(x)^2)(k)=\frac{1}{(2 \pi)^2}
\end{displaymath}
\begin{displaymath}
\times \int d^4 x \, e^{+ikx} \frac{1}{(2 \pi)^2} \int d^4 k'
\end{displaymath}
\begin{equation}
\hat{\Delta}_0^+ (k') e^{-ik'x}
\frac{1}{(2 \pi)^2}\int d^4k''
\hat{\Delta}_0^+ (k'') e^{-ik''x} \, ,
\end{equation}
and using
\begin{equation}
\int d^4 x \, e^{i(k-k'-k'')x}=(2 \pi)^4 \delta^{(4)}(k-k'-k'')
\end{equation}
this implies
\begin{displaymath}
\hat{\Delta}_2^+(k) = \frac{1}{(2 \pi)^2} \int d^4 k' \, d^4 k'' \, \hat{\Delta}_0^+ (k') \hat{\Delta}_0^+ (k'') \,
\delta^{(4)}(k-k'-k'')
\end{displaymath}
\begin{displaymath}
=\frac{1}{(2 \pi)^2} \int d^4 k' \,  \hat{\Delta}_0^+ (k') \hat{\Delta}_0^+ (k-k')
\end{displaymath}
\begin{equation}
=-(2 \pi)^{-4} \int d^4 k' \, \Theta(k'^0) \delta(k'^2) \Theta(k^0-k'^0) \delta((k-k')^2) \, . \label{convo}
\end{equation}
The integral eq. (\ref{convo}) vanishes if $k^0 \not \in \bar{V}^+$, i.e. if $k$ is not in the closed forward light-cone
\begin{equation}
 \bar{V}^+ = \{k | k^0 \ge 0, k^2 \ge 0 \}.
\end{equation}
In a Lorentz system where $k=(k^0>0, \vec{0})$, due to the first $\Theta-$ and
$\delta$-distribution in eq. (\ref{convo}) one has $E=|\vec{k}'|=k'_0$ and
\begin{equation}
\delta((k-k')^2) =\delta((k_0-{k'}_0)^2-E^2)=\delta(k_0^2-2k_0 E) \, ,
\end{equation}
therefore
\begin{displaymath}
\hat{\Delta}_2^+(k)= -(2 \pi)^{-4} \int \frac{d^3 k'}{2E} \,  \Theta(k^0-E) \delta(k_0^2-2 k^0 E)
\end{displaymath}
\begin{displaymath}
=-(2 \pi)^{-4} \int d|\vec{k'}| \,  \frac{4 \pi |\vec{k'}|^2}{2 |\vec{k'}|} \, \Theta(k^0-|\vec{k'}|)
\frac{\delta(k^0/2-|\vec{k'}|)}{|2 k^0|}
\end{displaymath}
\begin{equation}
\overset{|\vec{k'}|=k^0/2}{=}
-\frac{1}{4 (2 \pi)^3} \Theta(k^0) \, .
\end{equation}
For arbitrary $k$ follows
\begin{equation}
\hat{\Delta}_2^+(k) = -\frac{1}{4 (2 \pi)^3} \Theta(k^0) \Theta(k^2) \, .
\end{equation}

\noindent As a further step, the meaning of the expression $\Delta_n^+(x)=(\Delta_0^+(x))^n$
is investigated in configuration space. Obviously,
$\Delta_n^+(x)=(\Delta_0^+(x))^n \sim 1/x^{2n} = 1/(x^2)^n$ is very 'singular' in x-space.
For $n \ge 2$ and $x^2 \neq 0$ one easily derives
\begin{equation}
\Box \frac{1}{((x^0-i \varepsilon)^2 -\vec{x}^2)^n}=\frac{4n(n-1)}{((x^0-i \varepsilon)^2-\vec{x}^2)^{n+1}}
\end{equation}
translating into
\begin{equation}
\Box \Delta_n^+(x) = -16 i \pi^2  n (n-1) \Delta_{n+1}^+ (x) \,.
\end{equation}
In momentum space, this implies
\begin{equation}
-k^2 \hat{\Delta}_n^+(k) = -16 i \pi^2  n (n-1) \hat{\Delta}_{n+1}^+ (k)
\end{equation}
or
\begin{equation}
 \hat{\Delta}_{n+1}^+(k) = \frac{k^2}{16 i \pi^2  n (n-1)} \hat{\Delta}_{n}^+ (k) \, ,
\end{equation}
and inductively it follows for $n \ge 2$
\begin{equation}
\hat{\Delta}_n^+(k)= \frac{(-i)^n (k^2)^{n-2}}{4^{n-1}(2 \pi)^{2n-1} (n-1)!(n-2)!} \Theta(k^0) \Theta(k^2) \, .
\end{equation}
Hence, the Fourier transform of
\begin{displaymath}
D^+(x-y)=\langle 0 | : e^{-ig \varphi(x)} : : e^{+ig \varphi(y)} : | 0 \rangle
\end{displaymath}
\begin{equation}
 = \sum \limits_{n=0}^{\infty}
\frac{i^n}{n!} (g^2)^n [\Delta^+_0(x-y)]^n
\end{equation}
sums up to
\begin{displaymath}
\hat{D}^+(k)=(2 \pi)^2 \delta^{(4)}(k) + \frac{g^2}{2 \pi} \Theta(k^0) \delta(k^2)
\end{displaymath}
\begin{displaymath}
+\frac{g^4}{2! 4 (2 \pi)^3} \Theta(k^0) \Theta(k^2)+ \ldots =
\end{displaymath}
\begin{displaymath}
(2 \pi)^2 \delta^{(4)}(k) + \frac{g^2}{2 \pi} \Theta(k^0) \delta(k^2)
\end{displaymath}
\begin{equation}
+\sum \limits_{n=2}^{\infty} \frac{2(g^2)^n (k^2)^{n-2}}{(4 \pi)^{2n-1} n! (n-1)!(n-2)!} \Theta(k^0) \Theta(k^2) \, .
\end{equation}
This expression is, up to a normalization constant, the correct expression for eq. (14) in
\cite{Schroer_Concept}, where the combinatorial coefficients are stated incorrectly without a derivation.\\

\noindent In order to highlight the high-energy behaviour of the above expression we introduce the function
\begin{equation}
d(x)=\sum \limits_{n=2}^{\infty} \frac{x^n}{n! (n-1)! (n-2)!} \label{newf}
\end{equation}
For $x \gg 1$, $d(x)$ asymptotically behaves like
\begin{equation}
d(x) \sim \frac{1}{2 \pi \sqrt{3}} x^{2/3} e^{3 x^{1/3}} \, . \label{asymptotic_approx}
\end{equation}
The derivation of this result is given in Appendix B.
$\hat{D}^+(k)$ grows faster than any polynomial on the momentum-space forward light-cone.
Therefore, $\hat{D}^+$ does not belong to the Schwartz
space of tempered distributions, since an integral of the form
\begin{equation}
\hat{D}^+ (\hat{g}) = \int d^4 k \, \hat{D}^+(k) \hat{g}(k) \label{jaffe}
\end{equation}
does not exist for all $g$, $\hat{g} \in \mathcal{S}(\mathds{R}^4)$, despite the rapid decrease of
such functions. However,
the integral eq. (\ref{jaffe}) exists if $\hat{g}$ is of compact support. Unfortunately, a non-vanishing Fourier
transform $\hat{g}(k)$ implies that $g(x)$ does not have a compact support in configuration space, which
hampers the definition of causality according to eq. (\ref{causality}).\\

\noindent However, Jaffe \cite{Jaffe} has shown that it is still possible to construct a restricted space of
test functions in configuration space which contains test functions of compact support, such that the principle of
causality can be formulated and the fields in the derivative coupling model can be considered
operator valued distributions on the appropriately chosen test function space;
it is possible to find test functions with compact support which have
a Fourier transform decreasing so fast that integral like the one in eq. (\ref{jaffe}) exist. One finally may conclude
that even a physically trivial interaction may enforce a formalism which goes beyond the well-behaved
setting of Schwartz distributions, which lies at the basis of perturbatively renormalizable QFTs.

\subsection{Operator field equations of motion}
Eq. (\ref{deriv1}) contains the product of two field operators. A 'subtraction' or regularization
is necessary to define the equations of motion of the derivative coupling model. In fact,
normal ordering in the sense of a subtraction leads to
\begin{displaymath}
(\partial_\mu \varphi \gamma^\mu \psi)_{reg}(x)=:\partial_\mu \varphi(x) \gamma^\mu \psi(x):
\end{displaymath}
\begin{displaymath}
=\lim_{y \rightarrow x} \big[ \partial_\mu \varphi(x)  \gamma^\mu \psi(y) -
\langle 0 | \partial_\mu \varphi(x)  \gamma^\mu \psi(y)  | 0 \rangle \big]
\end{displaymath}
\begin{displaymath}
=\lim_{y \rightarrow x}
\big[ :\partial_\mu \varphi(x): : \gamma^\mu \psi_0(x) e^{-ig \varphi(x)}:
\end{displaymath}
\begin{equation}
- g \partial_\mu^x \Delta^+(x-y)
\gamma^\mu : \psi_0(x) e^{-ig \varphi(x)}: \big] \, .
\end{equation}

\section{'Fermionic' version of the derivative coupling model}
\subsection{Gauge charge operator for free fields}
Before turning back to the derivative coupling model, some remarks concerning the gauge structure of
perturbative quantum electrodynamics (QED) are in order.
Considering the free massless neutral vector potential prominent in QED obeying the wave equation
$\sq A^\mu(x)=0$ in Feynman gauge, the Fourier representation reads
($\omega=k^0=| \vec{k} |=\sqrt{k_1^2+k_2^2+k_3^2}$)
\begin{equation}
A^\mu (x)=\pdh\io\Bigl(a^\mu (\vec k)e^{-ikx}+a^\mu (\vec k)^\dagger
e^{ikx}\Bigl),\label{photon_decomp}
\end{equation}
and can be quantized in Lorentz-invariant form according to
\begin{equation}
[A^\mu (x),\, A^\nu (y)]=-i g^{\mu\nu} \Delta_0(x-y).\label{photon_commutator}
\end{equation}
The commutators of the absorption and emission parts alone are
\begin{equation}
[A^\mu _-(x),\, A^\nu _+(y)]=-i g^{\mu\nu} \Delta^+_0(x-y) \, , \label{photon_commutator1}
\end{equation}
\begin{equation}
[A^\mu _+(x),\, A^\nu _-(y)]=-i g^{\mu\nu} \Delta^-_0(x-y) \, .\label{photon_commutator2}
\end{equation}

\noindent In classical electrodynamics the vector potential can be changed by a gauge transformation
\begin{equation}
A'^\mu(x)=A^\mu(x)+\lambda\d^\mu u(x) \, , \label{gauge_op}
\end{equation}
with $u(x)$ again fulfilling the wave equation
$\sq u(x)=0$ since the transformed field $A'^\mu(x)$ still should satisfy the original wave
equation and the same
commutation relations eq. (\ref{photon_commutator}) as $A^\mu(x)$. This is true if the gauge
transformation eq. (\ref{gauge_op}) is of the following form
\begin{equation}
A'^\mu(x)=e^{-i\lambda Q}A^\mu(x)e^{i\lambda Q},\label{Lie_trafo}
\end{equation}
where $Q$ is some operator in the Fock-Hilbert space the photon field lives in. Expanding
eq. (\ref{Lie_trafo}) by means
of the Lie series
\begin{equation}
=A^\mu(x)-i\lambda [Q,A^\mu(x)]+O(\lambda^2)
\end{equation}
and a comparison with eq. (\ref{gauge_op}) leads to the condition
\begin{equation}
[Q,A^\mu(x)]=i\d^\mu u(x).\label{basic_comm}
\end{equation}

\noindent The operator $Q$ will be called gauge charge because it is the
infinitesimal generator of the gauge transformation defined by eq. (\ref{gauge_op}).
Its importance relies on the fact that the factor space given by the kernel and the closure
of the range of the gauge operator
$\mathcal{F}_{\rm ph}={\rm Ker}\, Q/\overline{{\rm{Ran}}\, Q}$ is isomorphic to the subspace of physical photon
states \cite{Krahe}, \cite{Razumov}. Before clarifying what this means, the following remarks are in order.\\

\noindent Firstly, it is not clear at the present status of the discussion
whether the field introduced in eq. (\ref{gauge_op}) has to be considered as
a classical C-number field or a quantum field. It will turn out that it can be treated as a classical
or a quantized bosonic field in QED, however, for non-abelian gauge theories like quantum chromodynamics
(QCD) the $u$-field necessarily becomes a fermionic scalar field, also called a ghost field. We will call the
massless scalar field $u$
a ghost field in the following irrespective of the fact whether it is quantized or not, bosonic or fermionic.
\\

\noindent Secondly, the commutator given in eq. (\ref{photon_commutator})
generates a problem for $\mu=\nu=0$: $g^{00}$ has the wrong sign, making it impossible to have
time-like photon states with positive norm if one insists on the hermiticity of the $A^0$-field component.
The positive frequency Pauli-Jordan distribution for time-like photons would acquire the opposite sign
as exhibited by eq. (\ref{positivity2}).
The situation is remedied by defining a so-called Krein structure \cite{Razumov}, \cite{Bognar}
on the photonic Fock-Hilbert space.
Introducing a conjugation $K$
\begin{equation}
a_0(\vec{k})^K = -a_0(\vec{k})^\dagger ,
\quad a_j(\vec{k})^K=a_j(\vec{k})^\dagger , \quad j=1,2,3 \, ,
\end{equation}
so that $A_\mu^K=A_\mu$, allows to maintain the positive-definiteness  on the Fock-Hilbert space
which is comprised in the definition of a Hilbert space,
however, the redefined field
\begin{equation}
A^0 (x)=\pdh\io\Bigl(a^0 (\vec k)e^{-ikx}-a^0 (\vec k)^\dagger
e^{ikx}\Bigl) = A^0_K \label{photon_decomp_redev}
\end{equation}
which will be used from now on is no longer a hermitian field.
In accordance with the commutation relations eq. (\ref{photon_commutator}) holds
\begin{equation}
[a^\mu(\vec{k}),a^\nu(\vec{k}')^\dagger]=\delta^{\mu \nu} \delta^{(3)}(\vec{k}-\vec{k}') \, ,
\end{equation}
\begin{equation}
[a^\mu (\vec{k}),a^\nu (\vec{k}')]=[a^\mu (\vec{k})^\dagger,
a^\nu (\vec{k}')^\dagger]=0 \, .
\end{equation}
Fortunately, abandoning the hermiticity of the zeroth component of the gauge potential  does
not invalidate the unitarity of the $S$-matrix in
QED on the physical subspace of transverse photons \cite{Krahe}.\\

\noindent The gauge transformation operator with the properties required so far turns out to be
\begin{equation}
Q=\int \limits_{x_0=const.} d^3 x \, \partial_\mu A^\mu(x)
\partial^{\! \! \! \! ^{^\leftrightarrow}}_0 u(x)  \, . \label{charge}
\end{equation}
$Q$ has the physical dimension of a scalar or vector field, an energy, or an inverse length squared.
It is sufficient for the moment to consider $u$ as a real C-number field.
In any case anticipated so far it can be shown that Q is a well-defined operator on the Fock space.
It is not important over which spacelike plane the integral
in eq. (\ref{charge}) is taken, since $Q$ is time independent:
\begin{displaymath}
\dot{Q}=\int \limits_{x_0=const.} d^3 x \,
(-\partial_0^2 \partial_\mu A^\mu u + \partial_\mu A^\mu
\partial_0^2 u)
\end{displaymath}
\begin{equation}
=\int \limits_{x_0=const.} d^3 x \,
(-\bigtriangleup \partial_\mu A^\mu u + \partial_\mu A^\mu
\bigtriangleup u)=0 \, .
\end{equation}
This formal proof uses the wave equation and partial
integration.
Another way to understand the time independence of the gauge
charge is to define the {\em{gauge current}}
\begin{equation}
j_g^\mu=\partial_\nu A^\nu  \! \stackrel
{\leftrightarrow}{\partial^\mu}  \! u, \quad Q=\int d^3x \, j^0_g \, ,
\end{equation}
which is conserved
\begin{equation}
\partial_\mu j_g^\mu=\partial_\mu (\partial_\nu A^\nu \partial^\mu
u-\partial^\mu \partial_\nu A^\nu u)=0 \, .
\end{equation}
Besides the crucial property of the gauge charge expressed by the
commutator with $A^\mu$
\begin{equation}
[Q,A^\mu(x)]=i \partial^\mu u(x) \, , \label{gaugecommutation}
\end{equation}
all higher commutators like
\begin{equation}
[Q,[Q,A^\mu(x)]]=0
\end{equation}
vanish for a bosonic or C-number ghost field $u$, but not for a fermionic ghost field.
Eq. (\ref{gaugecommutation}) can be derived by using some distributional properties
of the massless Pauli-Jordan distribution
\begin{equation}
\Delta^0(x)=-\frac{i}{(2 \pi)^3} \int d^4 k \,
\delta(k^2) \mbox{sgn}(k_0) e^{-i kx} \, .
\end{equation}
Using the identity
\begin{equation}
\delta(k^2)=\delta(k_0^2-\vec{k}^{\, 2})=\frac{1}{2 |k^0|}
\Bigl( \delta(k^0-|\vec{k}|) + \delta(k^0+|\vec{k}|) \Bigr) \, ,
\end{equation}
leads to
\begin{displaymath}
\partial_0 \Delta^0(x)= -\frac{i}{(2 \pi)^3} \int
\frac{d^4 k}{2 |k^0|}
\end{displaymath}
\begin{displaymath}
\Bigl( \delta(k^0-|\vec{k}|) - \delta(k^0+|\vec{k}| \Bigr)
(-i k^0) e^{-ikx}
\end{displaymath}
\begin{equation}
=-\frac{1}{2 (2 \pi)^{3}} \int d^3 k \,
\Bigl( e^{-i(|\vec{k}|x^0 -\vec{k} \vec{x})} +
e^{-i(-|\vec{k}|x^0 -\vec{k} \vec{x})} \Bigr) \, .
\end{equation}
Restricting this result to $x^0=0$ implies
\begin{equation}
\partial_0 \Delta^0(x) |_{x^0=0} =-(2 \pi)^{-3} \int d^3 k \,
e^{+i\vec{k} \vec{x}} = -\delta^{(3)} (\vec{x}) \, . \label{timederiv}
\end{equation}
In a completely analogous way, one derives for the derivatives of the Pauli-Jordan distribution
restricted to the space-like plane $x^0=0$
\begin{equation}
\partial_0^2 \Delta^0(x) |_{x^0=0} = 0 \, , \quad {\vec{\nabla}}
\Delta^0(x) |_{x^0=0}=0 \, . \label{deriv2j}
\end{equation}
Note that we always consider the well-defined differentiated
distribution first, which then gets restricted to a subset
of its support.
The commutator is now given explicitly by
\begin{displaymath}
[Q, A_\mu(y)]= [\int \limits_{x^0=y^0} d^3 x \,
\partial_\nu A^\nu (x) \stackrel
{\leftrightarrow}{\partial_0^x} u(x), A_\mu(y)]
\end{displaymath}
\begin{equation}
=-i\int \limits_{x^0=y^0} d^3 x \, \partial_\mu^x \Delta^0(x-y)
\stackrel {\leftrightarrow}{\partial_0^x} u(x) \, .
\end{equation}
Here, use was made of the freedom to choose any constant value
for $x^0$. Setting $x^0=y^0$, such that
$x^0-y^0=0$ and applying eqns. (\ref{timederiv}) and (\ref{deriv2j})
in the sequel,
one has for $\mu =0$
\begin{displaymath}
[Q, A_0(y)]=
-i\int \limits_{x^0=y^0} d^3 x \, \partial_0^x \Delta^0(x-y)
\stackrel {\leftrightarrow}{\partial_0^x} u(x)
\end{displaymath}
\begin{equation}
=i \int \limits_{x^0=y^0} d^3 x \, \delta^{(3)} (\vec{x}-\vec{y})
\partial_0^x u(x) = i \partial_0 u(y)
\end{equation}
due to the double timelike derivative of
$\Delta^0$ vanishing on the integration domain according to eq. (\ref{deriv2j}).
The result for the commutator of $Q$ with the spacelike
components of $A^\mu$ is also obtained by using eqns.
(\ref{timederiv}) and (\ref{deriv2j})
and by shifting the gradient acting of the Pauli-Jordan distribution
by partial integration on the ghost field.\\

\noindent From the Lie series
\begin{displaymath}
e^{-i \lambda Q} A^\mu e^{+i \lambda Q}=
A^\mu-\frac{i\lambda}{1 !} [Q,A^\mu]- \frac{\lambda^2}{2 !}
[Q,[Q,A^\mu]]+...
\end{displaymath}
\begin{equation}
=A^\mu-i \lambda [Q,A^\mu]=A^\mu+\lambda \partial^\mu u \, ,
\label{gaugeseries}
\end{equation}
follows that $Q$ is indeed a generator of gauge transformations for
a C-number ghost field $u$; it is a simple task to show that also
$[Q,i \partial^\mu u]=[Q,[Q, A^\mu]]=0$ holds in the case of a bosonic massless ghost field.\\

\noindent As a further step fermionic ghost fields are introduced.
$u(x)$ is assumed to be a fermionic scalar field with mass zero which has the
following Fourier decomposition ($\omega(\vec{k})=|\vec{k}|$)
\begin{equation}
u(x)=\pdh\int{d^3 k \over\sqrt{2 \omega(\vec{k})}}\Bigl(c_2(\vec k)e^{-ikx}+
c_1^\dagger (\vec k) e^{ikx}\Bigl) \, ,\label{ufield1}
\end{equation}
and in addition, a further scalar field shall be defined by
\begin{equation}
\tilde u(x)=\pdh\int{d^3 k\over\sqrt{2 \omega(\vec{k})}}\Bigl(-c_1(\vec k)e^{-ikx}+
c_2^\dagger (\vec k) e^{ikx}\Bigl) \, ,\label{def_u}
\end{equation}
with absorption and emission operators
$c_j$, $c_k^\dagger$ obeying the anticommutation relations
\begin{equation}
\{c_j(\vec k),c_k^\dagger (\vec k') \}=\delta_{jk}\delta^{(3)}(\vec k-\vec k') \, .
\label{def_u_tilde}
\end{equation}
Conventionally, the $\tilde{u}$-field is called an anti-ghost field.
The absorption and emission parts with the adjoint operators will be
indexed by $\pm$-signs below again. They satisfy the following anticommutation relations
\begin{equation}
\{u^-(x),\tilde u^+(y)\}=(2\pi)^{-3}\int {d^3k \over 2 \omega(\vec{k})}\,e^{-ik(x-y)}=
i \Delta^{+}(x-y) \, ,
\end{equation}
\begin{equation}
\{u^+(x),\tilde u^-(y)\}=-(2\pi)^{-3}\int {d^3k \over 2  \omega(\vec{k})}\,e^{ik(x-y)}=
i \Delta^{-}(x-y) \, . \label{commu_g}
\end{equation}
All other anticommutators vanish. This implies
\begin{equation}
\{u(x),\tilde u(y)\}=i \Delta(x-y)\label{ufield2}
\end{equation}
and $\{u(x), u(y)\}=0$.
Still the nilpotent gauge charge $Q$
satisfying eq. (\ref{basic_comm}) is given by
\begin{equation}
Q=\int d^3x\,[\d_\nu A^\nu\d_0u-(\d_0\d_\nu A^\nu)u]
\=d\int d^3x\,\d_\nu A^\nu{\dl}_0u \label{Q_def_1}
\end{equation}
where the integrals are taken over any plane $x^0={\rm const.}$\\

\noindent Using the Leibnitz rule $\{AB,C\}=A\{B,C\}-[A,C]B$  for graded algebras for the present
gauge charge for massless spin-1 fields
\begin{displaymath}
Q^2=\frac{1}{2} \{Q,Q\} =\frac{1}{2} \int \limits_{x_0=const.}
d^3 x \partial_\nu A^\nu(x) \{  \stackrel
{\leftrightarrow}{\partial}_{0} u(x) , Q \}
\end{displaymath}
\begin{equation}
-\frac{1}{2}  \int \limits_{x_0=const.}
d^3 x [\partial_\nu A^\nu(x) , Q] {\dl}_{0}
u(x) = 0
\end{equation}
together with the facts that $\{ u(x), u(y) \}=0$ and
\begin{equation}
[\partial_\nu A^\nu(x) , Q]= -i \partial_\nu \partial^\nu u(x) = 0
\end{equation}
finally shows that $Q$ is nilpotent.\\

\noindent On the ghost sector, the Krein structure is introduced by
\begin{equation}
c_2^K (\vec k) = c_1^\dagger (\vec k),\quad c_1^K (\vec k)=
c_2^\dagger (\vec k) \, ,
\end{equation}
so that $u^K=u$ is $K$-selfadjoint and $\tilde u^K=-\tilde u$. Then
$Q$ is densely defined on the Fock-Hilbert space
and becomes $K$-symmetric $Q
\subset Q^K$. Roughly speaking, the K-conjugation is the
natural generalization of the usual hermitian conjugation
to the full (unphysical) Fock space $\mathcal{F}$ which contains time-like and longitudinal photons as
well as the fermionic ghost states. Again, positivity on the Fock-Hilbert space can only be maintained
by the introduction of the Krein structure. Enforcing $K=\dagger$ would necessitate the existence
of negative norm states in the ghost sector. The strategy preferred here is based on a true
Hilbert space approach.\\

\noindent It is convenient to introduce bosonic operators which destroy or create unphysical
photon states which are a combination of time-like and longitudinal states
\begin{equation}
b_{1,2}=(a_{\|} \pm a_0)/\sqrt{2} \,  , \quad
a_{\|}=k_j a^j/|\vec{k}|\, ,
\end{equation}
satisfying ordinary commutation relations
\begin{equation}
[b_i (\vec{k}), b^\dagger_j (\vec{k}) ] = \delta_{ij} \delta^{(3)} (\vec{k} -\vec{k}') \, .
\end{equation}

\noindent Then, the gauge charge $Q$ itself is given by
\begin{equation}
Q = \sqrt{2} \int d^3 k \, \omega(\vec{k}) [ b_2^\dagger ( \vec{k}) c_2 ( \vec{k} ) +
c_1^\dagger (\vec{k}) b_1 (\vec{k}) ] \, . \label{momentum_gauge}
\end{equation}
The explicit form of the gauge charge reveals that it generates a transformation where
unphysical photon states are transformed into ghost states and vice versa.
The transverse physical photon states remain unaffected by a gauge transformation.\\

\noindent A calculation using the decomposition of the anticommutator
\begin{displaymath}
\{ b_1^\dagger (\vec{k}) c_1 (\vec{k}), c_1^\dagger (\vec{k}') b_1 (\vec{k}') \}=
\end{displaymath}
\begin{displaymath}
b_1^\dagger ( \vec{k}) \{ c_1 ( \vec{k}), c_1^\dagger (\vec{k}') \} b_1 (\vec{k}') -
[b_1^\dagger ( \vec{k} ) , c_1^\dagger (\vec{k}') b_1 (\vec{k}') ] c_1 (\vec{k})=
\end{displaymath}
\begin{equation}
(b_1^\dagger (\vec{k}) b_1(\vec{k}) + c_1^\dagger (\vec{k}) c_1 (\vec{k})) \delta^{(3)} (\vec{k}-\vec{k}')
\end{equation}
shows that the anticommutator
\begin{displaymath}
\{Q^\dagger,Q\} = 2 \int d^3 k \, \vec{k}^2
\end{displaymath}
\begin{equation}
\bigl[ b_1^\dagger(\vec{k}) b_1(\vec{k}) +
b_2^\dagger(\vec{k})b_2(\vec{k})+c_1^\dagger(\vec{k})c_1(\vec{k})+
c_2^\dagger(\vec{k})
c_2(\vec{k}) \bigr]  \label{anti}
\end{equation}
 is essentially the number operator for unphysical particles
apart from the phase space factor $\omega(\vec{k})^2=\vec{k}^2$.
Therefore, if a state $| \Phi \rangle $ in the Fock-Hilbert space satisfies $\{Q^\dagger, Q \} |\Phi \rangle = 0$,
it contains physical transverse photon states only. Hence, the physical Hilbert space is the kernel
\begin{equation}
\mathcal{F}_{phys} = \rm{Ker} \{ Q^\dagger , Q \} \, . \label{specification}
\end{equation}
Additionally, since $\{ Q^\dagger , Q \} = Q^\dagger Q + Q Q^\dagger$ is self-adjoint and positive
\begin{equation}
\langle \Phi | (Q^\dagger Q + Q Q ^\dagger) \Phi \rangle = || Q \Psi ||^2 + || Q^\dagger \Phi ||^2 \ge 0 \, .
\label{positive_QQ}
\end{equation}
This expression vanishes only iff $Q \Phi = Q^\dagger \Phi = 0$, leading to another characterization
of the physical Hilbert space
\begin{equation}
\mathcal{F}_{phys} = \rm{Ker} \,  Q \cap \rm{Ker} \, Q^\dagger \, .
\end{equation}
$\rm{Ker} \, Q$ is a subspace of $\mathcal{F}$ and orthogonal to the closure $\overline{\rm{Ran} \, Q^\dagger}$
of the range of $Q^\dagger$, since for $| \Phi \rangle \in \rm{Ker} \, Q$ one has
\begin{equation}
\langle Q \Phi | \Psi \rangle = 0 =  \langle \Phi | Q^\dagger \Psi \rangle \, .
\end{equation}
In fact, $\mathcal{F}$ has the direct decomposition
\begin{equation}
\mathcal{F} = \rm{Ker} \, Q \oplus \overline{\rm{Ran} \, Q^\dagger} = \rm{Ker} \, Q^\dagger \oplus
\overline{{\rm{Ran} \, Q}} \, .
\end{equation}
This can be proven by noticing that the domain $\rm{Dom} (Q^\dagger)$ is dense in $\mathcal{F}$, so if
$\langle \Upsilon | Q^\dagger \Psi \rangle =0$ for all $\Psi \in \rm{Dom} (Q^\dagger )$, then $\langle Q \Upsilon
| \Psi \rangle =0$, implying $ | Q \Upsilon \rangle =0$ or $| \Upsilon \rangle  \in \rm{Ker} \, Q$.
Using the nilpotency $Q^2=0$ one sees from $\langle Q^\dagger \Psi | Q \Phi \rangle =
\langle \Psi | Q^2 \Phi \rangle
0=$ that $\overline{\rm{Ran} \, Q^\dagger}$ is orthogonal to $\overline{\rm{Ran} \, Q}$. Consequently,
$\mathcal{F}$ has the direct decomposition
\begin{equation}
\mathcal{F} =
\overline{\rm{Ran} \, Q^\dagger}\oplus \overline{\rm{Ran} \, Q} \oplus \mathcal{F}_{phys} \, .
\end{equation}
Indeed, if $P_1$ and $P_2$ are projection operators on the first two subspaces above,
due to orthogonality one has $P_1 P_2 = 0 = P_2 P_1$. It follows that the projection
operator on the orthogonal complement of $P_{1,2}$ is given by
\begin{equation}
1-(P_1+P_2) = (1-P_1) ( 1-P_2) \,  ,
\end{equation}
which is the projection onto $\rm{Ker} \,  Q \cap \rm{Ker} \, Q^\dagger$, the physical subspace.
Obviously,
\begin{equation}
\rm{Ker} \, Q = \mathcal{F}_{phys} \oplus \overline{\rm{Ran} \, Q} \, ,
\end{equation}
accordingly
\begin{equation}
\mathcal{F}_{phys} = \rm{Ker} \, Q / \overline{\rm{Ran} \, Q} \, .
\end{equation}
One may note that $\rm{Ran} \, Q = \rm{Dom} (Q^{-1})$ is indeed not closed since $Q^{-1}$ is unbounded
for a massless gauge field $A^\mu$.\\

\noindent Returning to the defining property of $Q$ as being the infinitesimal
generator of gauge transformations given by eq. (\ref{Lie_trafo}) and eq. (\ref{basic_comm}),
the notation
\begin{equation}
d_QF=[Q,F] \, ,
\end{equation}
if the (normally ordered) product of free fields
$F$ contains only bosonic fields and an even number of ghost fields,
and
\begin{equation}
d_QF=\{Q, F\}=QF+FQ \, , \label{odd_ghost}
\end{equation}
if $F$ contain an odd number of ghost fields, may be introduced for practical reasons. Then $d_Q$ has all
properties of an anti-derivation, in particular the identity
\begin{equation}
\{AB,C\}=A\{B,C\}-[A,C]B\label{anti_Leibniz}
\end{equation}
implies the product rule
\begin{equation}
d_Q(F(x)G(y))=(d_QF(x))G(y)+(-1)^{n_F}F(x)d_QG(y) \, , \label{product_rule}
\end{equation}
where $n_F$ is the ghost number of $F$, i.e. the number of $u$'s in $F$ minus
the number of $\tilde u$-fields. The gauge variations $d_Q$ of some free fields
now are
\begin{equation}
d_Q A^\mu=i\d^\mu u,\quad d_QA_\pm^\mu=i\d^\mu u_\pm \, , d_Q u=0 \, ,
 \label{another_label}
\end{equation}
\begin{equation}
d_Q\tilde u=\{Q,\tilde u\}=-i\d_\mu A^\mu,\quad
d_Q\tilde u_\pm=-i\d_\mu A_\pm^\mu \, .\label{dQ_label}
\end{equation}
The latter follows from the anticommutation relation eq. (\ref{ufield2}).
$d_Q$ changes the ghost number by one, i.e. a bosonic field
goes over into a fermionic field and vice versa. Then the nilpotency $Q^2=0$
implies for a bosonic field $F_B$
\begin{displaymath}
d_Q^2F_B=\{Q,[Q,F_B]\}
\end{displaymath}
\begin{equation}
=Q(QF_B-F_BQ)+(QF_B-F_BQ)Q=0 \, ,
\end{equation}
and for a Fermi field $F$
\begin{displaymath}
d_Q^2F=[Q,\{Q,F\}]
\end{displaymath}
\begin{equation}
=Q(QF_B+F_BQ)-(QF_B-F_BQ)Q=0 \, ,
\end{equation}
hence
\begin{equation}
d_Q^2=0
\end{equation}
is also nilpotent. For such situations one can use notions from homological algebra, for example, if
\begin{equation}
F = d_Q G \, ,
\end{equation}
the $F$ is called a coboundary \cite{Massey}.
The gauge variation $d_Q$ has some similarity with the
BRST transformation in the functional approach to QCD. However, the BRST transformation
operates on interacting fields (mainly classical) and the quantum gauge
invariance which will be defined below for free fields displays some technical differences compared to
BRST invariance \cite{Kugo}.\\

\noindent To end this section, the operator gauge transformation when working with fermionic ghosts shall
be considered. It is straightforward to see that the Lie series
\begin{equation}
e^{-i \lambda Q} A^\mu e^{+i \lambda Q}=
A^\mu-\frac{i\lambda}{1 !} [Q,A^\mu]- \frac{\lambda^2}{2 !}
[Q,[Q,A^\mu]]+...
\end{equation}
terminates after the second order term. Since
\begin{displaymath}
[Q, u(x)]= Q u(x) - u(x) Q = \{Q , u(x) \} - 2 u(x) Q
\end{displaymath}
\begin{equation}
= -2 u(x) Q \, ,
\end{equation}
one has
\begin{equation}
[Q, u(x)Q]=[Q,u(x)] Q + u(x)[Q,Q] = 0 \, ,
\end{equation}
or, stated equivalently
\begin{displaymath}
[Q, u(x)Q]=Qu(x)Q-u(x)Q^2
\end{displaymath}
\begin{equation}
= Qu(x)Q+u(x) Q^2=\{ Q,u(x) \} Q = 0 \, .
\end{equation}
Consequently, the gauge transformation of the gauge potential is found to be given by
\begin{displaymath}
A'^\mu (x) = A^\mu(x) + \lambda \partial^\mu u(x) + i \lambda^2 \partial^\mu u(x) Q
\end{displaymath}
\begin{equation}
= A^\mu (x) + \partial^\mu u(x) ( \lambda+ i \lambda^2 Q) \, . \label{gauge_trafo_A}
\end{equation}
Analogously, one finds for the ghost fields
\begin{equation}
u'(x)=u(x)+2 i \lambda u(x) Q \, ,
\end{equation}
\begin{equation}
\tilde{u}'(x) = \tilde{u}(x) + \lambda ( 2i \tilde{u} (x) Q - \partial_\mu A^\mu (x)) -i \lambda^2
\partial_\mu A^\mu (x) Q \, .
\end{equation}

\subsection{Definition of perturbative quantum gauge invariance}
We take the next step towards full QED and couple photons to
electrons.
In perturbative QED, the $S$-matrix is expanded as a power series
in the coupling constant $e$. At first order, the interaction
is described by the normally ordered product of free fields
\begin{equation}
{\cal{H}}_{int}(x) = -{\cal{L}}_{int}^{_{QED}}(x) = -\mbox{e}:\bar{\Psi}(x)
\gamma^\mu \Psi(x): A_\mu (x) \, , \label{foc}
\end{equation}
where $\Psi$ is the electron field operator and $\mbox{e} = -e >0$ the elementary charge.
The $S$-matrix is then usually given in the literature
by the formal expression ($T$
denotes time ordering)
\begin{displaymath}
S={\bf{1}}
\end{displaymath}
\begin{displaymath}
+\sum \limits_{n=1}^{\infty} \frac{(-i)^n}{n !}
\int  \limits_{\mathds{R}^{4n}} d^4 x_1 \ldots d^4 x_n \, T[{\cal{H}}_{int}(x_1) \ldots
{\cal{H}}_{int}(x_n)]
\end{displaymath}
\begin{equation}
={\bf{1}}+\sum \limits_{n=1}^{\infty} \frac{1}{n !}
\int d^4 x_1 \ldots d^4 x_n \, T_n(x_1, \ldots x_n), \label{smatrix}
\end{equation}
where we have introduced the time-ordered products $T_n$
for notational simplicity, and we have
\begin{equation}
T_1(x)=-i {\cal{H}}_{int}(x)=i \mbox{e} :\bar{\Psi}(x)
\gamma^\mu \Psi(x): A_\mu (x). \label{first_order_QED}
\end{equation}
Expression (\ref{smatrix}) is plagued by
infrared and ultraviolet divergences.
We leave this technical problem aside
and we assume that the $T_n$ are already regularized, well-defined
operator valued distributions, which are symmetric in the
space coordinates $(x_1, \ldots x_n)$.\\

\noindent A precise definition of perturbative quantum gauge invariance for QED, which works
in a very analogous way for QCD, can be derived by investigating
how infinitesimal gauge transformations act on the higher orders of the perturbative $S$-matrix.
One considers the (anti-)commutators
\begin{displaymath}
[Q,A_\mu]=i\partial_\mu u  ,
\quad \{Q,u\}=0  , \quad \{Q,\tilde{u}\}=-i\partial_\nu A^\nu  \, ,
\end{displaymath}
\begin{equation}
\quad [Q,\Psi]=[Q,{\bar{\Psi}}]=0 \, . \label{commu_f}
\end{equation}
The commutators of $Q$ with the electron field are of course trivial,
since the operators act on different Fock space sectors.
Only the first and the last two commutators in eq. (\ref{commu_f}) are needed here,
the others would become important in QCD. Note, however, that ordinary commutation
relations of the electron field with $Q$ or the ghost fields $u$ and $\tilde{u}$ can be switched into
anticommutation relations by a Klein transformation (see \cite{Araki} and references
therein) without changing the physical content of the theory.
From eq. (\ref{gaugeseries}) one knows that the commutator of $Q$ with an
operator gives the first order variation of the operator subject to a
gauge transformation.
Then, for the first order interaction $T_1$
\begin{displaymath}
[Q,T_1(x)] = - \mbox{e} :\bar{\Psi} (x) \gamma^\mu \Psi (x) : \partial_\mu u(x)
\end{displaymath}
\begin{equation}
=i \partial_\mu (i \mbox{e} :\bar{\Psi} (x) \gamma^\mu \Psi (x) :u(x))
= i \partial_\mu T^\mu_{1/1} (x) \, . \label{gi}
\end{equation}
Here, electron current conservation was used
\begin{equation}
\partial_\mu :\bar{\Psi} \gamma^\mu \Psi:=0 \, .
\end{equation}
Note that the \emph{free} electron field is {\em{not}} affected by the
gauge transformation.
The term
\begin{equation}
T^\mu_{1/1}=i \mbox{e} :\bar{\Psi} \gamma^\mu \Psi:u \, ,
\end{equation}
called
the 'Q-vertex' or 'gauge vertex' of QED, can be used in a generalized manner
from the first order eq. (\ref{gi}) to $n$-th order
\begin{displaymath}
[Q,T_n(x_1,...x_n)] = i \sum_{l=1}^{n} \partial_\mu^{x_l} T_{n/l}^{\mu}
(x_1,...x_n)
\end{displaymath}
\begin{equation}
= (\hbox{\rm sum of divergences}) \, , \label{dive}
\end{equation}
where $T^\mu_{n/l}$ is again a mathematically well-defined version of
the time-ordered product
\begin{equation}
T^\mu_{n/l}(x_1,...,x_n) \,
\, \mbox{'} \!= \! \mbox{'} \, \, T(T_1(x_1)...T^\mu_{1/1} (x_l)...T_1(x_n)) \, ,
\end{equation}
thereby defining by eq. (\ref{dive}) the condition of gauge invariance in QED
\cite{HKS}.\\

\noindent If one considers for a fixed $x_l$ all terms in $T_n$
with the external field operator $A_\mu(x_l)$
\begin{equation}
T_n(x_1,...x_n) = :t^\mu_l(x_1,...x_n) A_\mu(x_l):+...
\end{equation}
(the dots represent terms without $A_\mu(x_l)$),
then gauge invariance eq. (\ref{dive}) requires
\begin{equation}
\partial_\mu^l [t^\mu_l(x_1,...x_n)u(x_l)]=t^\mu_l(x_1,...x_n) \partial_\mu
u(x_l)
\end{equation}
or
\begin{equation}
\partial_\mu^l t^\mu_l(x_1,...x_n) = 0 \, , {\label{eich}}
\end{equation}
i.e. one obtains the Ward-Takahashi identities \cite{Ward} for QED.
The Ward-Takahashi identities express the implications of
gauge invariance of QED, which is defined here on the
operator level, by C-number identities for Green's distributions.\\

\noindent The main important property of gauge invariance of perturbative QED can be stated as follows:
There exists a symmetry transformation generated by the gauge charge $Q$, which
leaves the $S$-matrix elements invariant, since the gauge transformation
only adds divergences in the analytic sense
to the $S$-matrix expansion which vanish after integration
over the coordinates $x_1, \dots x_n$.\\

\noindent The observation that QED is gauge invariant is interesting on its own,
but the true importance of gauge invariance is the fact that it allows to prove on a formal level
the {\em{unitarity}} of the $S$-matrix on the physical subspace
(see the last paper of \cite{HKS}).
Due to the presence of the skew-adjoint operator $A^0$ in the first order coupling
term eq. (\ref{foc}) which defines the interaction between fermions and gauge fields
or the related presence of unphysical ghost and
longitudinal and timelike photon states in QED formulated in a local and renormalizable gauge,
the $S$-matrix is not unitary on the full Fock space, but it is on ${\cal{F}}_{phys}$.
An full algebraic proof shall not be given here, but we emphasize that
gauge invariance is the basic prerequisite which ensures
unitarity, a fact which becomes plausible when one assures oneself that a
gauge transformation acts only on the unphysical sector of a gauge theory.
A detailed discussion of this fact can be found in
\cite{Razumov}, \cite{HKS}, \cite{pgi}.
Ghosts are introduced only as a formal but convenient tool,
they 'blow up' the Fock space and they do not
interact with the electrons and photons. In QCD, the situation
is far more complicated than in QED when non-perturbative aspects of the theory have to be
considered.\\

\noindent The perturbative expression eq. (\ref{smatrix}) is problematic,
because the time-ordered products $T_n$ are operator valued distributions after regularization,
and they have to be smeared out by test functions. In order to be more precise in a
mathematical sense,  one has to  introduce a test function
$g_0(x) \in {\cal{S}}({\bf{R}}^4)$ normalized such that $g_0(0)=1$ and replace
expression (\ref{smatrix}) by
\begin{displaymath}
S={\bf{1}}
\end{displaymath}
\begin{equation}
+\sum \limits_{n=1}^{\infty} \frac{1}{n !}
\int d^4 x_1 \ldots d^4 x_n \, T_n(x_1, \ldots x_n)
g_0(x_1)...g_0(x_n).
\end{equation}
Here, $g_0$ acts as an infrared regulator, which switches
off the long range part of the interaction in theories where
massless fields are involved.
E.g., in QED the emission of soft photons is switched
by $g_0$, and as long as the so-called
adiabatic limit
$g_0 \rightarrow 1$ has not been performed, $S$-matrix elements
remain finite.
One possibility to perform the adiabatic limit is by
scaling the switching function $g_0(x)$, i.e. one replaces
$g_0(x)$ by $g(x)=g_0(\epsilon x)$ and performs the
limit $\epsilon \rightarrow 0$, such that $g$ and the coupling strength everywhere approaches
a constant value.
If the $S$-matrix is modified by a gauge transformation, operators which are divergences are
added to the n-th order term $T_n$. Such a contribution can be written as
\begin{displaymath}
\int d^4 x_1 ... d^4 x_n \,
\end{displaymath}
\begin{displaymath}
\partial_\mu^{x_l} O^{... \mu... }(x_1,...,x_l,...x_n) g(x_1) ...
g(x_l) ...g(x_n)
\end{displaymath}
\begin{displaymath}
=-\int d^4 x_1 ... d^4 x_n \,
\end{displaymath}
\begin{equation}
 O^{... \mu... }(x_1,...,x_l,...x_n) g(x_1) ...
\partial_\mu^{x_l} g(x_l) ...g(x_n). \label{farout}
\end{equation}
In the adiabatic limit, the gradient $\partial_\mu^{x_l} g(x_l)$
vanishes. Unfortunately, this property of the scaling limit
does not guarantee that the whole term eq. (\ref{farout}) vanishes.
Introducing a switching function $g_0$ is the natural infrared regularization
in the framework of operator valued distributions, but it destroys the
Poin\-car\'e invariance of the theory and leads to a problem to define
the physical vacuum. Whereas this problem more or less might be under control
for QED, it is a serious problem expressed by the catchwords 'infrared slavery' for
QCD. The infrared problem is not
really understood in QCD, and all proofs of unitarity
which exist in the literature have to be taken with a grain
of salt, because they are avoiding the discussion of infrared
problems somehow.\\

\noindent The fermionic derivative coupling model defined in the following section emerges as
a special limit when one considers perturbative QED with a vanishing coupling constant $e$, maintaining only
an unphysical part of the interaction.

\subsection{The model}
Starting from the field equations again, keeping in mind that one has to take care of the order of products
in the case of fermionic fields, one has
\begin{equation}
(i \gamma_\mu \partial^\mu -m ) \psi(x) = g \partial^\mu \varphi(x) \gamma_\mu \psi(x) \,  , \label{deriv1f}
\end{equation}
\begin{equation}
\Box \varphi(x) = 0 \, , \label{deriv2f}
\end{equation}
\begin{equation}
\Box A^\mu(x) = 0 \, . \label{deriv3f}
\end{equation}
The gauge field $A^\mu(x)$ is rather an additional spectator.
The coupling term in eq. (\ref{deriv1f}) emerges when considering a gauge transformed version of
the first order coupling term in QED given by eq. (\ref{first_order_QED}) according to eq. (\ref{gauge_trafo_A}),
in the limit where $e \rightarrow 0$ but $e \lambda^2=g$ held fixed.\\

\noindent An operator solution of the equation of motion above reads, defining
$\varphi(x)=-i Qu(x)$ by
the help of the gauge charge operator given in eq. (\ref{momentum_gauge}) and the fermionic scalar
field with the properties displayed by eqns. (\ref{ufield1}) - (\ref{ufield2}),
\begin{displaymath}
\psi(x)= e^{-ig\varphi(x)} \psi_0(x)
\end{displaymath}
\begin{displaymath}
=[1-g Qu(x)+g^2 Qu(x)Q u(x) + \ldots] \psi_0(x)
\end{displaymath}
\begin{displaymath}
=[1-g Qu(x)-g^2 Q^2u(x) u(x)-\ldots] \psi_0(x)
\end{displaymath}
\begin{equation}
= [1-ig \varphi(x)] \psi_0 (x) \, , \label{op_sol}
\end{equation}
using the free fields $A^\mu(x)$ and $\psi_0(x)$ acting on the Fock-Hilbert space introduced in the
discussion of QED,
satisfying
\begin{equation}
\Box A^\mu(x) = 0, \quad (i \gamma_\mu \partial^\mu -m ) \psi_0(x) =0 \, ,
\end{equation}
and $\varphi(x)$ satisfying the commutation relation
\begin{displaymath}
[\varphi(x),\varphi(y)]= - [Q u(x), Q u(y)]
\end{displaymath}
\begin{displaymath}
=-Q u(x) Q u(y) + Qu(y) Q u(x)
\end{displaymath}
\begin{equation}
=Q^2 u(x) u(y) - Q^2 u(y) u(x) = 0 \, .
\end{equation}
Since $\{Q,u\}=0$, we have $Qu=-uQ$. Additionally, $Q$ is nilpotent $Q^2=0$.
$u$ is an unphysical Fermi field, $u(x)u(x)= \, :u(x) u(x): \, = - :u(x) u(x): \, =0$ and
$:\partial^\mu u(x) u(x):  \, =0$ holds and similar identities hold for $\varphi$,
accordingly
\begin{equation}
\psi_0(x) =  [1+ig \varphi(x)][1-ig \varphi(x)] \psi_0 (x) = [1+ig \varphi(x)] \psi (x) \, .
\end{equation}
Inserting the operator solution eq. (\ref{op_sol}) into eq. (\ref{deriv1f}) leads to
\begin{displaymath}
(i \gamma_\mu \partial^\mu-m) \psi(x)=
i \gamma_\mu \partial^\mu ([1-ig \varphi(x)] \psi_0 (x))-m \psi (x)=
\end{displaymath}
\begin{displaymath}
[1-ig \varphi(x)] [i \gamma_\mu  \partial^\mu \psi_0 (x)-m \psi_0(x)] + g \partial^\mu \varphi (x)
\gamma_\mu  \psi_0 (x)=
\end{displaymath}
\begin{equation}
g \partial^\mu \varphi (x)   [1+ig \varphi(x)]  \gamma_\mu  \psi (x) =
g   \partial^\mu \varphi (x)  \gamma_\mu \psi (x) \, .
\end{equation}
The interaction term is unphysical and gauge invariant in the sense that
\begin{equation}
[Q, {\cal{H}}_{int}(x)]=-g [Q,:\bar{\psi}(x) \gamma^\mu \psi(x): \partial_\mu \varphi(x)]=0 \, .
\end{equation}
$\mathcal{H}_{int}$ is K-symmetric like $\varphi^K =(-i Q u)^K=i u^K Q^K = -i Q u=\varphi$.\\

\noindent The model presented above can be modified in the following way. 
Let $a(x)$ be a C-number field with
$a(0,\vec{x}) \in \mathcal{S}(\mathds{R}^3)$ satisfying the wave equation
$\Box a(x)=0$.  Then one has the Fourier decompositions
\begin{equation}
a(x)= \int \frac{d^3k}{\sqrt{2 (2 \pi)^3 \omega(\vec{k})}}
\Bigl( a_- (\vec{k}) e^{-ikx} +a_+(\vec{k}) e^{+ikx} \Bigr) \, ,
\end{equation}
\begin{equation}
\partial_0 a(x)= i \int d^3k \sqrt{\frac{\omega(\vec{k})}{2(2 \pi)^3}}
\Bigl( -a_- (\vec{k}) e^{-ikx} +a_+(\vec{k}) e^{+ikx} \Bigr) \, ,
\end{equation}
again with $k^0=\omega(\vec{k})=|\vec{k}|$ and
$kx=k^0 x^0 - \vec{k} \vec{x}$
and analogous Fourier representations hold for the operator valued distributions $u(x)$ and
$\partial_0 u(x)$.\\

\noindent The definition of the operator
\begin{equation}
\tilde{Q}= \int \limits_{x_0=const.} d^3 x \, a(x) \partial^{\! \! \! \! ^{^\leftrightarrow}}_0 u(x)
\label{newcharge}
\end{equation}
is time-independent, for $x_0=0$ one obtains
\begin{displaymath}
\tilde{Q}= \frac{i}{(2 \pi)^3} \int d^3 x \int \frac{d^3 k'}{\sqrt{2 \omega(\vec{k}')}} \int d^3 k
\sqrt{\frac{\omega(\vec{k})}{2}}
\end{displaymath}
\begin{displaymath}
\Bigl[ \Bigl( a_- (\vec{k}') e^{i\vec{k}' \vec{x}} +a_+(\vec{k}') e^{-i\vec{k}' \vec{x}} \Bigr)
\Bigl( -c_2 (\vec{k}) e^{i\vec{k} \vec{x}} +c_1^\dagger (\vec{k}) e^{-i\vec{k} \vec{x}} \Bigr)
\end{displaymath}
\begin{displaymath}
-\Bigl(-a_- (\vec{k}) e^{i\vec{k} \vec{x}} +a_+(\vec{k}) e^{-i\vec{k} \vec{x}} \Bigr)
\Bigl( c_2 (\vec{k}') e^{i\vec{k}' \vec{x}} +c_1^\dagger (\vec{k}') e^{-i\vec{k}' \vec{x}} \Bigr)
\Bigr]
\end{displaymath}
\begin{displaymath}
= \frac{i}{2} \int d^3 k \Bigl[ -a_-(-\vec{k}) c_2(\vec{k}) - a_+ (\vec{k}) c_2 (\vec{k})
\end{displaymath}
\begin{displaymath}
+ a_-(-\vec{k}) c_2(\vec{k}) - a_+ (\vec{k}) c_2 (\vec{k})
\end{displaymath}
\begin{displaymath}
+ a_-(\vec{k}) c_1^\dagger(\vec{k}) + a_+ (-\vec{k}) c_1^\dagger (\vec{k})
\end{displaymath}
\begin{displaymath}
+  a_-(\vec{k}) c_1^\dagger(\vec{k}) - a_+ (-\vec{k}) c_1^\dagger (\vec{k}) \Bigr]
\end{displaymath}
\begin{equation}
= i \int d^3 k [-a_+(\vec{k}) c_2(\vec{k}) + a_-(\vec{k}) c_1^\dagger(\vec{k}) ] \, .
\end{equation}
Again one has $\tilde{Q}^2=\frac{1}{2} \{ \tilde{Q}, \tilde{Q} \} =0$, therefore the model discussed
above can be formulated with $\tilde{Q}$ instead of $Q$ without a quantized vector field $A^\mu$
when $a^*_-(\vec{k})=a_+(\vec{k})$ is invoked, i.e. $a(x)$ must be real. Then $Q$ becomes $K-$symmetric,
since
\begin{equation}
Q^K = i \int d^3 k  [-a^*_-(\vec{k}) c_2(\vec{k}) + a^*_+(\vec{k}) c_1^\dagger(\vec{k}) ] 
\end{equation}
and the Krein correlator of the $\psi-$field remains trivial
\begin{equation}
\langle 0 | \psi_0(x) \bar{\psi}_0(y) | 0 \rangle =
\langle 0 | \psi_0(x) \psi^K_0(y) | 0 \rangle
= \langle 0 | \psi(x) \psi^K (y) | 0 \rangle \, .
\end{equation}
However, since
\begin{equation}
\{ \tilde{Q}^\dagger, Q \}= \int d^3 k ( |a_-(\vec{k})|^2 + |a_+ (\vec{k})|^2 ) \, ,
\end{equation}
the original specification of the physical space according to eq. (\ref{specification}) is lost.
It is left to the reader to couple the ghost field $u$ instead of $\varphi$ to $\psi$ in the
same way as a simple exercise.\\

\noindent The fermionic model is physically trivial, the formalism rather involved, but also one possible variant
of the classical derivative coupling model which served here for the introduction of concept related to the operator
gauge formalism. Non-re\-nor\-mali\-zable expressions or non-tempered distributions nowhere appear,
despite the dimension of the coupling term.

\section*{Conclusions}
The two models presented in this work are a tool to demonstrate the fact that
there are several ways to quantize a classical field theory. The models
also clarify that the r\^ole of fields is rather to implement the principle of causality, but the type
and number of the fields appearing in a theory is rather unrelated to the physical spectrum
of empirically observable particles. The fields are coordinatizations of an underlying physical
theory and carriers of charges which finally serve to extract the algebra of observables.\\

\noindent From a distributional point of view, theories based on point-like localized quantum
fields may indicate that the frame of Schwartz operator valued distributions favoured in
perturbative QFT is too narrow, but it remains unclear whether a loss of the original concepts
using tempered distributions can be avoided within a suitable formalism.

\section{Appendix A: A distributive toolbox}
\subsection{Support}
A distribution $d \in \mathcal{S}'(\mathds{R}^n)$ is called regular, if it can be represented by
\begin{equation}
d(f)=\int \limits_{\mathds{R}^n} dx \, d(x) f(x) \, ,
\end{equation}
where $d(x)$ is a locally integrable function and $f \in \mathcal{S}(\mathds{R}^n)$.
This close analogy between functions and distributions leads to the definition of the support of distributions.
\noindent The support of a function defined on  $\mathds{R}^n$ is the closure of the set where the function
is non-zero
\begin{equation}
\mbox{supp}(f) = \overline{\{ x \in \mathds{R}^n | \, f(x) \neq 0 \}} \, .
\end{equation}

\noindent A point $x$ belongs to the support of a distribution $d$ iff for every neighbourhood $U_x$ of $x$
a function $f$ exists with $\mbox{supp}(f) \subset U_x$ and  $d(f) \neq 0$.

\subsection{Tensor product of distributions}
Let $d_1 \in \mathcal{S}'(\mathds{R}^n)$,  $d_2 \in \mathcal{S}'(\mathds{R}^m)$.
Then a unique distribution $h \in \mathcal{S}'(\mathds{R}^{n+m})$ exists such that
for all $f_1(x) \in \mathcal{S}(\mathds{R}^n),$ $ f_2(y) \in \mathcal{S}(\mathds{R}^m)$

\begin{equation}
h(f_1(x) f_2(y)) = d_1(f_1(x)) d_2(f_2(y))  \, .
\end{equation}
$h=d_1 \otimes d_2$ is the \emph{tensor product} of $d_1$ and $d_2$.
A simple example is given by the product of Dirac distributions
\begin{equation}
\delta^{(n)} (x) = \delta(x^1) \delta(x^2) \ldots \delta(x^n) \, , \quad x=(x^1,x^2, \ldots x^n) \, ,
\end{equation}
where
\begin{equation}
\int \limits_{\mathds{R}^n}  d^n x \, \delta^{(n)}(x) f(x) = f(0) \, .
\end{equation}
The Fourier transform of the above distribution is given by
\begin{displaymath}
\hat{\delta}^{(n)} (k) = (2 \pi)^{-n/2} \, ,
\end{displaymath}
\begin{equation}
\int \limits_{\mathds{R}^n} d^n x \, e^{i k^1 x^1 + \ldots + i k^n x^n} = (2 \pi)^n
\delta^{(n)} (k) \, .
\end{equation}
In close analogy, tensor products of free fields , e.g., the product of two scalar fields on $\mathds{R}^4$
like $\varphi(x) \varphi(y)$ are again operator valued distributions, in the present case on $\mathds{R}^8$.
However, products like $\delta (x) \delta (x)$ (or $\varphi(x) \varphi(x)$) are ill-defined, but can
be regularized (by normal ordering) in order to define well-defined (operator-valued) distributions.

\subsection{Principal values and regularization}
\noindent An important distribution is $P \frac{1}{x}$, i.e. the principal value of  the singular
function $1/x \in C(\mathds{R} \backslash 0)$ interpreted as a distribution:
\begin{equation}
P \frac{1}{x}(f) = \lim \limits_{\varepsilon \searrow 0} \int \limits_{|x| > \varepsilon} dx \,
\frac{f(x)}{x}  \, , \quad
P \frac{1}{x} = \frac{d}{dx} \ln |x| \, .
\end{equation}

\noindent $P\frac{1}{x}$ is a \emph{regularization} of the divergent expression $\frac{1}{x}$.
Without regularization, $1/x$ is only defined on
\begin{equation}
\mathcal{S}_0(\mathds{R}) = \{f \in \mathcal{S}(\mathds{R}) \, | \, f(0)=0\} \, ,
\end{equation}
where the singular behaviour
of $1/x$ at $x=0$ gets absorbed.
$P\frac{1}{x}$ can be viewed as an extension of $\frac{1}{x} \Bigl|_{\mathcal{S}_0(\mathds{R})}$
to the whole test function space $\mathcal{S}(\mathds{R})$ according to the \emph{Hahn-Banach} theorem.
One may also write
\begin{equation}
P\frac{1}{x}(f)= \int \limits_{0}^{\infty} dx \, \frac{f(x)-f(-x)}{x} \, .
\end{equation}

\noindent A canonical regularization of the divergent, non-regularized integral
\begin{equation}
d_{1/x^2}^{\, nr} (f) =  \int \limits_{\mathds{R}} dx \, \frac{f(x)}{x^2}
\end{equation}
is possible by shifting a derivative
\begin{equation}
d_{1/x^2} (f)= \int \limits_{\mathds{R}} dx \, P \frac{1}{x} f'(x) \, .
\end{equation}

\noindent Equivalently, one may regularize
\begin{equation}
(x^{-2},f)_{reg}= \int \limits_{0}^{\infty} dx \,  \frac{f(x)+f(-x)-2f(0)}{x^2} \, .
\end{equation}

\subsection{Renormalization}
In regularization procedures, a distribution declared by a divergent expression becomes properly redefined
within a range of permissible solutions allowed by physical conditions. Subsequent renormalizations within
this range then may be performed.
It is often exploited that certain distributions exhibit a specific
scaling behaviour. E.g., the renormalization
\begin{equation}
d_{1/x^2} \rightarrow d_{1/x^2} + C \cdot \delta'(x)
\end{equation}
respects the scaling behaviour ($\lambda >0$) of the distribution $d_{1/x^2}$, because
\begin{displaymath}
\delta'(f) = -f'(0) \overset{formally}{=} \int  \limits_{\mathds{R}}
dx \, \delta'(x) f(x)
\end{displaymath}
\begin{equation}
= -\int \limits_{\mathds{R}} dx  \, \delta(x) f'(x)
\end{equation}
scales as
\begin{displaymath}
\int \limits_{\mathds{R}} dx \,  \delta' (\lambda x) f(x) \overset{x'= \lambda x}{=} \int \limits_{\mathds{R}}
\frac{dx'}{\lambda} \delta'(x') f(x'/\lambda)
\end{displaymath}
\begin{equation}
=-\int \limits_{\mathds{R}} \frac{dx'}{\lambda^2}  \delta(x') f'(x'/\lambda)= -\frac{1}{\lambda^2} f'(0) \, ,
\end{equation}
i.e.
\begin{equation}
\delta'(\lambda x) = \lambda^{-2} \delta'(x)
\end{equation}
 and
\begin{equation}
d_{1/x^2}(\lambda x)=(\lambda x)^{-2}_{reg}
=\lambda^{-2} d_{1/x^2}(x) \, .
\end{equation}

\subsection{Sokhotsky-Plemelj formula}
The distributions
\begin{equation}
\frac{1}{x \pm i0}= P \frac{1}{x} \mp i \pi \delta(x) \, ,
\end{equation}
are often constructed from a limiting procedure
\begin{equation}
\int \limits_{\mathds{R}} \frac{f(x)}{x+i0} \, dx = \lim \limits_{\varepsilon \searrow 0}
\int \limits_{\mathds{R}} \frac{f(x)}{x+i \varepsilon} \, dx \, .
\end{equation}
One easily derives the distributive identities below by considering the logarithm in the complex plane
where $\log(z)=\log|z|+i \mbox{Arg}(z)$
\begin{displaymath}
\frac{d}{dx} \log(x+i0)= \frac{1}{x+i0} = \frac{d}{dx} \log(|x|) + \frac{d}{dx} ( i  \pi \Theta(-x))
\end{displaymath}
\begin{equation}
= P \frac{1}{x} - i \pi
\delta(x)  \, .
\end{equation}
Differentiating $n$ times leads to
\begin{displaymath}
\frac{d}{dx} \frac{1}{x+i \epsilon} =- \frac{1}{(x+i \epsilon)^2} \, , \, \,
\frac{d^2}{dx^2} \frac{1}{x+i \epsilon} =+ \frac{2}{(x+i \epsilon)^3} \, , \, \, \ldots
\end{displaymath}
\begin{equation}
\frac{d^n}{dx^n} \frac{1}{x+i \epsilon} =(-1)^n \frac{n!}{(x+i \epsilon)^{n+1}} \, ,
\end{equation}
therefore
\begin{equation}
\frac{1}{(x+i0)^{n+1}}= P \frac{1}{x^{n+1}}-(-1)^{n} \frac{i \pi}{n!} \delta^{\{n\}}(x) \, ,
\end{equation}
where $\delta^{\{n\}}(x)$ denotes the $n$-fold derivative of $\delta(x)$ here, not the
$n-$dimensional Dirac distribution often used in the paper.

\subsection{An important remark}
A multiplication of tempered distributions which is commutative and associative can not be defined in general.
One has
\begin{equation}
(x \delta(x)) P \frac{1}{x} = 0 P \frac{1}{x} = 0 \neq \delta(x) (x P \frac{1}{x}) = \delta(x) \, .
\end{equation}
Unfortunately, distribution theory is linear. This is the origin of ultraviolet divergences in perturbative
QFT. The problem may be illustrated by an analogy where one considers
the Heaviside-$\Theta$- and Dirac-$\delta$-distributions in 1-dimensional
'configuration space'.
The product of these two distributions $\Theta(x) \delta(x)$ is
obviously ill-defined, however, the distributional Fourier transforms
\begin{equation}
\sqrt{2 \pi} \mathcal{F} \{ \delta \} (k)=
\sqrt{2 \pi} \hat{\delta} (k) = \int  \limits_{\mathds{R}} dx \, \delta(x) e^{-ikx} =1,
\end{equation}
\begin{displaymath}
\sqrt{2 \pi} \hat{\Theta} (k) = \lim_{\epsilon \searrow 0}
\int  \limits_{\mathds{R}} dx \, \Theta(x) e^{-ikx-\epsilon x}
\end{displaymath}
\begin{equation}
= \lim_{\epsilon \searrow 0}
\frac{ie^{-ikx-\epsilon x}}{k-i \epsilon} \Biggr|^{\infty}_{0}= -\frac{i}{k-i0},
\end{equation}
exist and one may attempt to calculate the ill-defined product in 'momentum space',
which \emph{formally} goes over into a convolution
\begin{displaymath}
\sqrt{2 \pi} \mathcal{F} \{ \Theta \delta \} (k)=\int  \limits_{\mathds{R}} dx \, e^{-ikx} \Theta(x)  \delta(x)
\end{displaymath}
\begin{equation}
=\int  \limits_{\mathds{R}} dx \, e^{-ikx} \int \limits_{\mathds{R}} \frac{dk'}{\sqrt{2 \pi}}
\hat{\Theta}(k') e^{+ik'x}
\int  \limits_{\mathds{R}} \frac{dk''}{\sqrt{2 \pi}} \hat{\delta}(k'') e^{+ik''x}.
\end{equation}
Since $\int  \limits_{\mathds{R}} dx \, e^{i(k'+k''-k)x}=2 \pi \delta(k'+k''-k)$, one obtains
\begin{equation}
\sqrt{2 \pi} \mathcal{F} \{ \Theta \delta \} (k) = \int  \limits_{\mathds{R}} dk' \,
\hat{\Theta} (k') \hat{\delta} (k-k') = {
-\frac{i}{2 \pi} \int  \limits_{\mathds{R}} \frac{dk'}{k'-i0}}.
\end{equation}
The obvious problem in x-space leads to a 'logarithmic UV divergence' in k-space.
A concise description of the scaling properties
of distributions, related to the wide-spread notion of power counting and
the superficial degree of divergence of Feynman integrals, is crucial for the correct
treatment of singular products of distributions in perturbative QFT.
There, the r\^{o}le of the {Heaviside $\Theta$-distribution} is taken over
by the time-ordering operator. The well-known textbook expression for the
perturbative $S$-matrix given by
\begin{displaymath}
S  = \sum \limits_{n=0}^{\infty} \frac{(-i)^n}{n!}
\int \limits_{-\infty}^{+\infty} dt_1 \ldots
\int \limits_{-\infty}^{+\infty} dt_n \, {T}
[H_{int}(t_1) \ldots H_{int}(t_n)]
\end{displaymath}
\begin{equation}
 =  \sum \limits_{n=0}^{\infty} \frac{(-i)^n}{n!}
\int   \limits_{\mathds{R}^4} d^4 x_1 \ldots \int  \limits_{\mathds{R}^4}  d^4 x_n \, {T}
[\mathcal{H}_{int}(x_1) \ldots
\mathcal{H}_{int}(x_n)], \label{Smatrix_textbook}
\end{equation}
where the interaction Hamiltonian {$H_{int}(t)$}
is given by the interaction Hamiltonian density $\mathcal{H}_{int}(x)$ via
\begin{equation}
H_{int}(t)=\int d^3 x \,
\mathcal{H}_{int}(x) \, ,
\end{equation}
is problematic in the UV regime (and in the infrared
regime, when massless fields are involved).
A time-ordered expression \`a la
\begin{displaymath}
T [\mathcal{H}_{int}(x_1) \ldots \mathcal{H}_{int}(x_n)]
\end{displaymath}
\begin{displaymath}
= \! \!
\sum \limits_{Perm. \, \, \Pi} \Theta(x^0_{\Pi_1}-x^0_{\Pi_2}) \ldots
\Theta(x^0_{\Pi_{(n-1)}}-x^0_{\Pi_n})
\end{displaymath}
\begin{equation}
\times \mathcal{H}_{int}(x_{\Pi_1})
\ldots \mathcal{H}_{int}(x_{\Pi_n})
\end{equation}
is formal (i.e., ill-defined), since the operator-valued distribution
products of the $\mathcal{H}_{int}$ are simply too singular to be multiplied by
$\Theta$-distributions.

\section{Appendix B: Asymptotic behaviour of $d(x)$}
The symbol $\sim$ will be used in the following for asymptotic approximations, i.e.
$f(x) \sim \Phi(x)$ if $f(x)/\Phi(x)$ tends to unity for $\mathds{R} \ni x \rightarrow + \infty$ according to
Landau \cite{Landau}. Then $f$ is asymptotic to $\Phi$, or $\Phi$ is an asymptotic approximation to $f$.\\

\noindent From the well-known identities for (double or odd) factorials
\begin{equation}
(2n)! = (2n)!! (2n-1)!! \, , \quad (2n)!!=2^n n! 
\end{equation}
and
\begin{equation}
(2n-1)!!=\frac{2^n}{\sqrt{\pi}} \Gamma \biggl( n+\frac{1}{2} \biggr)= \frac{2^n}{\sqrt{\pi}}
\biggl( n-\frac{1}{2} \biggr) !
\end{equation}
one readily obtains from
\begin{equation}
\biggl( n-\frac{1}{2} \biggr) ! \sim \frac{n!}{\sqrt{n}}
\end{equation}
the asymptotic approximation
\begin{equation}
(2n)! \sim \frac{2^{2n}}{\sqrt{\pi}} \frac{(n!)^2}{\sqrt{n}} \, .
\end{equation}
Using Stirling's formula, this result can be generalized to
\begin{displaymath}
(3n)! \sim \sqrt{2 \pi (3n)} \biggl( \frac{3n}{e} \biggr)^{3n} =
\frac{\sqrt{3} 3^{3n}}{2 \pi n} \Biggl[ \sqrt{2 \pi n} \biggl( \frac{n}{e} \biggr)^n \Biggr]^3
\end{displaymath}
\begin{equation}
\sim \frac{\sqrt{3} 3^{3n}}{2 \pi n} (n!)^3 \, .
\end{equation}
Accordingly, $d(x)$ in eq. (\ref{newf}) can be approximated by
\begin{displaymath}
d(x)=\sum \limits_{n=2}^{\infty} \frac{x^n}{n! (n-1)! (n-2)!} =
\sum \limits_{n=2}^{\infty} \frac{n^2 (n-1)}{(n!)^3} x^n
\end{displaymath}
\begin{equation}
\sim \sum \limits_{n=0}^{\infty} \frac{n^3}{(n!)^3} x^n
\sim \frac{\sqrt{3}}{2 \pi} \sum \limits_{n=0}^{\infty} \frac{3^{3n} n^2}{(3n)!} x^n
\end{equation}
or
\begin{equation}
d(x) \sim  \frac{\sqrt{3}}{2 \pi} \sum \limits_{n=0}^{\infty} \frac{n^2}{(3n)!} \bigl( 3x^{1/3}
\bigr )^{3n} \, .
\end{equation}
A straightforward, but rather tedious calculation shows that
\begin{displaymath}
 \sum \limits_{n=0}^{\infty} \frac{n^2}{(3n)!} y^{3n} =
 \frac{1}{27} y (y+1) e^y
 \end{displaymath}
 \begin{equation}
  -\frac{y}{27} e^{-\frac{y}{2}} \Biggl[ (y+1) \cos \biggl( \frac{\sqrt{3}y}{2} \biggr) - \sqrt{3}(y-1)
  \sin \biggl( \frac{\sqrt{3} y}{2} \biggr) \Biggr] \, .
 \end{equation}
 Note that the sine term contains an additional factor $\sqrt{3}$ which is missing in the
 cosine term. Keeping only the dominant term,
 one has asymptotically
 \begin{equation}
 \sum \limits_{n=0}^{\infty} \frac{n^2}{(3n)!} y^{3n} \sim
 \frac{y^2}{27} e^y \, ,
 \end{equation}
and setting $y=3x^{1/3}$ leads to the desired result
\begin{equation}
d(x) \sim \frac{\sqrt{3}}{2 \pi} \frac{(3 x^{1/3})^2}{27} e^{3x^{1/3}}
= \frac{1}{2 \pi \sqrt{3}} x^{2/3} e^{3x^{1/3}} \, .
\end{equation}

\end{document}